\documentclass[fleqn,usenatbib]{mnras}

\usepackage{newtxtext,newtxmath}

\usepackage[T1]{fontenc}

\DeclareRobustCommand{\VAN}[3]{#2}
\let\VANthebibliography\thebibliography
\def\thebibliography{\DeclareRobustCommand{\VAN}[3]{##3}\VANthebibliography}

\usepackage{graphicx}	
\usepackage{amsmath}	
\usepackage{multirow}
\usepackage{subcaption}
\usepackage{ulem}

\title[Evolution of MBH Binaries in Collisionally Relaxed NSCs]{Evolution of Massive Black Hole Binaries in Collisionally Relaxed Nuclear Star Clusters - Impact of Mass Segregation}

\author[Mukherjee et al.]{
Diptajyoti Mukherjee,$^{1}$\thanks{E-mail: diptajym@andrew.cmu.edu (DM)}
Qirong Zhu,$^{1}$
Go Ogiya$^{2,3,4}$
Carl L. Rodriguez$^{1}$
and Hy Trac$^{1,5}$
\\
$^{1}$McWilliams Center for Cosmology,Department of Physics, Carnegie Mellon University, Pittsburgh, PA 15213, USA\\
$^{2}$Institute for Astronomy, 
School of Physics, Zhejiang University, 
Hangzhou 310027, China\\
$^{3}$Waterloo Centre for Astrophysics,
University of Waterloo, 
Waterloo, ON N2L 3G1, Canada\\
$^{4}$Department of Physics and Astronomy,
University of Waterloo, 
200 University Avenue West, 
Waterloo, Ontario N2L 3G1, Canada\\
$^{5}$NSF AI Planning Institute for Physics of the Future, Carnegie Mellon University, Pittsburgh, PA 15213, USA
}

\date{Accepted XXX. Received YYY; in original form ZZZ}

\pubyear{2022}

\begin{document}
\label{firstpage}
\pagerange{\pageref{firstpage}--\pageref{lastpage}}
\maketitle

\begin{abstract}
Massive Black Hole (MBH) binaries are considered to be one of the most important sources of Gravitational Waves (GW) that can be detected by GW detectors like LISA. However, there are a lot of uncertainties in the dynamics of MBH binaries in the stages leading up to the GW-emission phase. It has been recently suggested that Nuclear Star Clusters (NSCs) could provide a viable route to overcome the final parsec problem for MBH binaries at the center of galaxies. NSCs are collisional systems where the dynamics would be altered by the presence of a mass spectrum. In this study, we use a suite of high-resolution $N$-body simulations with over 1 million particles to understand how collisional relaxation under the presence of  a mass spectrum of NSC particles affects the dynamics of the MBH binary under the merger of two NSCs. We consider MBH binaries with different mass ratios and additional non-relaxed models. We find that mass-segregation driven by collisional relaxation can lead to accelerated hardening in lower mass ratio binaries but has the opposite effect in higher mass ratio binaries. Crucially, the relaxed models also demonstrate much lower eccentricities at binary formation and negligible growth during hardening stages leading to longer merger timescales.  
The results are robust and highlight the importance of collisional relaxation on changing the dynamics of the binary. Our models are state-of-the-art, use zero softening, and high enough particle numbers to model NSCs realistically.  

\end{abstract}


\graphicspath{{./}{figures/}}


\section{Introduction}

Observations in the past twenty years have demonstrated that Massive Black Holes (MBHs), despite being point sources in the center of galaxies, play a vital role in galaxy evolution and growth \citep[e.g.,][]{2001AIPC..586..363K, 2013ARA&A..51..511K}. In the hierarchical growth of structures, galaxies frequently merge and form larger systems \citep[e.g.,][]{RodriguezGomez2016MNRAS.458.2371R}. Upon the merger of galaxies, two MBHs get the opportunity to come close to one another to form a bound pair. Recent observation searches have revealed the presence of well-separated accreting MBHs seen as multiple Active Galactic Nuclei (AGN) in a single galaxy, as well as circumstantial evidence for bound Keplerian binaries \citep[e.g.,][]{2003ApJ...582L..15K, 2006MmSAI..77..733K, 2015ASSP...40..103B}. 

MBH binaries are purported to be one of the strongest sources of Gravitational Waves (GW) in the universe. Observations of GW during the final inspiral phase of the binary would reveal information not only regarding the merger history of the galaxy over time but also constrain dynamical properties of galactic nuclei surrounding the MBH binary. Therefore, modeling the dynamical evolution of MBH binaries inside galactic nuclei is central to the astrophysical interpretation of galactic environment and dynamics. MBH binaries are expected to be sources of millihertz (mHz) GW that will be detectable by future space based GW detectors like LISA \citep{2017arXiv170200786A} or Tianqin \citep{2016CQGra..33c5010L}.

Due to variances in galactic environments in which they are embedded in, the dynamics of MBH coalescence is up for much debate. It is theorized that the merger is a three step process before the the final GW-emission step \citep[][chapter 8]{Begelman1980Natur.287..307B,2013degn.book.....M}. In the first step, the dynamical friction of stars and dark matter and that of the interstellar gas play a role in reducing the angular momentum of the black holes which then sink towards the center of the merged galaxy. When the black holes get close enough, they form a bound binary which signals the beginning of the second stage. This stage proceeds rapidly and the separation of the binary decreases due to dynamical friction and three-body scattering events. The third stage prior to GW-driven coalescence begins as the black holes form a hard binary. Once that happens further orbital decay occurs via three-body scattering.
If not enough MBH binary-star scattering occurs, the orbital decay of the binary stalls before it can reach the GW-emission phase. This is called the final parsec problem \citep[e.g.,][]{Milo2003AIPC..686..201M}.

The timescale associated with the shrinkage of the binary in the hard binary stage is unclear and heavily depends on the environment. For example, in spherical gas-poor galaxies, simulations have shown that the orbital decay of the MBH binary essentially halts due to the lack of stars in the loss-cone and the GW merger timescales often exceed the Hubble time \citep[e.g.,][]{Vasilev2015ApJ...810...49V}. Additionally, in cosmological simulations MBH seeds have been found to be inefficient at sinking to the center of the nuclei leading to longer merger timescales \citep[e.g.,][]{Ma2021MNRAS.508.1973M}. In case of merging galaxies, however, the non-sphericity of the merger product introduces global torques which can populate the loss-cone more effectively leading to a continued orbital decay of the MBH binary system. Simulations have also shown that merger timescales in such cases are less than the Hubble time \citep[e.g.,][]{Berczik2006ApJ...642L..21B,Khan2013ApJ...773..100K, Vasilev2015ApJ...810...49V,Vasilev2017ApJ...848...10V}. 

One way to overcome the final parsec problem is by embedding the MBHs in Nuclear Star Clusters (NSCs). Nuclear Star Clusters (NSCs) represent some of the densest stellar systems in the universe. They can have mass densities of $\rho \geq 10^6 M_{\odot} \rm{pc}^{-3}$ \citep[e.g.,][]{2020A&ARv..28....4N}. As the name suggests, NSCs are found in galactic nuclei. The masses and presence of NSCs correlate with the mass of the host galaxy. \cite{2019ApJ...878...18S} showed that their presence in galaxies depends on the stellar mass of the host galaxies, with a peak of 90\% at $M_{\rm stellar} \sim 10^9 M_{\odot}$. NSCs and MBHs can coexist in many cases. In fact, the Milky Way galaxy contains an NSC at the Galactic Center that has an MBH embedded in it \citep[e.g.,][]{2008ApJ...689.1044G}. Using data from \cite{2019ApJ...878...18S}, \cite{2020MNRAS.493.3676O} speculate that under the assumption that all NSCs contain an MBH at the center, 50\% of all Milky Way sized galaxies should have both an NSC and and MBH present in their nuclei. 

In \cite{2020MNRAS.493.3676O}, the authors show that if MBHs are embedded in NSCs prior to merger, tidal effects from the merging NSCs accelerate the orbital evolution timescale before and around the time the binary is formed, thus circumventing the final parsec problem. In the presence of NSCs the formation of a hard binary occurs faster and the whole process of decay into the GW regime is accelerated. The authors found that the mergers were extremely efficient with lower mass ratio binaries merging in $\sim 100 $ Myr while binaries with mass ratio of unity merging in $~5$ Gyr, which is still less than the Hubble time. 

Since NSCs are collisional stellar systems they undergo collisional relaxation even under the presence of an MBH at the center. A collisionally relaxed state implies the presence of a Bahcall-Wolf cusp \citep[][]{1976ApJ...209..214B}. If there is a mass spectrum present, the more-massive objects form a steeper cusp than the less-massive objects \citep[][]{1977ApJ...216..883B, 2009ApJ...697.1861A}.  During the merger of two galaxies containing NSCs, in the absence of MBH binaries, the cusp is expected to be retained \citep{2005MNRAS.360..892D}. The presence of MBH binaries leads to a partial or complete disruption of the cusp \citep[e.g.,][]{2005MNRAS.360..892D}.

 In \cite{2020MNRAS.493.3676O}, the authors considered a one-component mass function to study the effect of NSCs on MBH binaries. Realistic NSCs, however, are comprised of a spectrum of masses \citep[e.g.,][]{Preto2010ApJ...708L..42P,2012ApJ...744...74G}. The effects of a mass spectrum have been explored previously by \cite{2012ApJ...744...74G} and \cite{Khan2018A&A...615A..71K} using a fixed binary mass ratio. However, a systematic comparison of unsegregated versus segregated models as a function of the binary mass ratio is missing. A segregated Bahcall-Wolf cusp could lead to enhanced hardening rates and could potentially accelerate the evolution to GW driven coalescence stage. In addition, the effects of the relaxed cusp on the hardening rates of the binary would be an interesting investigation. We, therefore, are motivated to understand how collisionally relaxed NSCs, under the presence of a mass spectrum, affect the dynamics of the binary in the stages leading up to GW driven coalescence.

In this work, we extend the models presented in \cite{2020MNRAS.493.3676O} to understand the effects of mass-segregation in NSCs on the dynamics of the MBH binary. With the usage of higher mass resolution compared to previous studies and a two-component mass function, our models are able to better represent realistic NSCs. We use the Fast Multipole Method (FMM) \citep{greengard1987fast,cheng1999fast} based $N$-body code {\tt\string Taichi} \citep{2021NewA...8501481Z,2021ApJ...916....9M} which has been shown to reproduce collisional effects as accurately as direct-summation based $N$-body codes while using a fraction of the computational time. To understand the effects of mass segregation we use a two-component mass function: one consisting of objects roughly a solar mass or less and the other consisting of heavier objects like stellar mass black holes. 
We systematically study the effect of relaxed, segregated cusps on the dynamics of MBH binaries with different mass ratios using a suite of $N$-body simulations and discuss the role of the relaxed cusp in the merger process. 

The paper is organized as follows: in section \ref{sec:methods} we describe the numerical methods and improvements made to {\tt\string Taichi} to handle dense systems more accurately. In section \ref{sec:models}, we describe the models of the mergers. In section \ref{sec:results} we provide the results and in section \ref{sec:discussion} we discuss the impact of stochasticity and compare our results to those of previous studies. This is followed conclusions section \ref{sec:conclusions}.

 \label{sec:intro}

\section{Numerical Methods} \label{sec:methods}

We perform a suite of $N$-body simulations using $N \sim 1.32\times 10^6 $ particles  to study the formation and evolution of the MBHs embedded in NSCs and their evolution after the formation of the MBH binary.  
The simulation setup is similar to that presented in \cite{2020MNRAS.493.3676O} with changes to improve the resolution and the modeling of the clusters to include a two-component mass species. The detailed description of the models is provided in the next section.

To simulate the system, we utilize the FMM based code {\tt\string Taichi} \citep{2021NewA...8501481Z, 2021ApJ...916....9M}. \cite{2021ApJ...916....9M} showed that {\tt\string Taichi} can simulate systems as accurately as direct-summation based collisional $N$-body codes while scaling as $\mathcal O(N)$. The accuracy of the force calculation in {\tt\string Taichi} can be tuned via the usage of an input accuracy parameter ($\epsilon$) which controls the opening angle and a multipole expansion parameter ($p$) which controls the number of expansion terms used in the force calculation. Using {\tt\string Taichi} we can simulate large-$N$ systems without the usage of specialized hardware within a physically reasonable amount of time. In this work, we extend {\tt\string Taichi} to include a fourth-order force-gradient integrator and regularization using the {\tt\string AR-Chain} scheme \citep{1999MNRAS.310..745M}. Additionally, we improve the accuracy of the force solver in {\tt\string Taichi} in this work. We briefly detail the improvements below.

\subsection{Updated Integration Scheme}

Contemporary direct-summation based $N$-body codes use a fourth order time integration scheme like the Hermite method \citep[e.g.,][]{Makino1992PASJ...44..141M}. In our previous work \citep{2021ApJ...916....9M} we adopted an integrator based on hierarchical Hamiltonian splitting which is only second-order accurate \citep{Pelupessy2012NewA...17..711P}. In this work, we extend {\tt\string Taichi} to include a novel fourth-order force-gradient integrator \citep{2021MNRAS.502.5546R} which decomposes the system into slow and fast subsystems based on the interaction timesteps of the particles. The fast subsystem is then hierarchically split until the slow-fast split results in no particles in the fast subsystem. We refer to this scheme as the {\tt\string HHS-FSI} scheme hereafter. Hamiltonian splitting integrators are suitable due to the large dynamical range present in our simulations. Unlike conventional composition symplectic integrators presented by Yoshida \citep{1990PhLA..150..262Y}, {\tt\string HHS-FSI} utilizes strictly positive sub-steps made possible by computing an additional gradient term along with the Newtonian accelerations \citep{1997PhLA..226..344C,Chen2005CeMDA..91..301C} which is only possible because the potential term does not depend on momentum.

Unlike \cite{2021MNRAS.502.5546R} where the gradient term is calculated by direct summation, we utilize an extrapolation method described in \cite{Omelyan2006PhRvE..74c6703O} which uses a fictitious middle step to approximate the gradient-force term. Tests by \cite{Omelyan2006PhRvE..74c6703O} have shown this approach is indistinguishable from the Chin \& Chen \citep{Chen2005CeMDA..91..301C} method.  To do this, we follow the method \citep[see also][section 3.1]{Farr2007ApJ...663.1420F} where
\begin{gather}
    \mathbf{p} \leftarrow \mathbf{p} - \frac{h}{6}\mathbf{D}V(\mathbf{q}) \\
    \mathbf{q} \leftarrow \mathbf{q} + \frac{h}{2}\frac{\mathbf{p}}{m} \\
    \mathbf{p} \leftarrow \mathbf{p} - \frac{2h}{3} \mathbf{D}V \left(\mathbf{q}-\frac{h^2}{24m}\mathbf{D}V(\mathbf{q})\right) \\
    \mathbf{q} \leftarrow \mathbf{q} + \frac{h}{2}\frac{\mathbf{p}}{m}\\
    \mathbf{p} \leftarrow \mathbf{p} - \frac{h}{6} \mathbf{D}V(\mathbf{q})
\end{gather}
where $h$ is the step-size, $\mathbf{q}$ is the position, $\mathbf{p}$ is the momentum and $\mathbf{D}V$ represents the gradient of the potential $V$ evaluated at the position given in the parenthesis. In total, four calls to the Poisson solver are needed in one step.

The forward symplectic nature ensures more accurate and efficient integration compared to that of the Yoshida scheme \citep[e.g.,][]{Chin2007PhRvE..75c6701C}. The integrator is manifestly momentum conserving and includes an individual symmetrized timestepping scheme similar to that described in \cite{2021ApJ...916....9M}. Despite the loss of symplecticity due to the usage of individual timesteps, the usage of time symmetrization ensures that there is no secular drift in the energy leading to much better energy conservation \citep[e.g.,][]{Makino2006NewA...12..124M}. 

\subsection{Algorithmic regularization}
Even with the inclusion of a fourth order scheme, treatment of close encounters with the MBH binary can prove to be computationally challenging. However, hierarchical Hamiltonian splitting integrators are easy to modify to include regularization as a result of clean separation of fast system from the slow system. This enables accurate handling of close binaries and/or addition of post-Newtonian terms as one can plug-in any accurate few-body solvers to evolve the Hamiltonian of the fast system. Therefore, in order to handle the dynamics of the MBH binary and its interactions with scattering particles more accurately, we include regularization in {\tt\string Taichi}. 

We have utilized the {\tt\string SpaceHub} API \citep{2021MNRAS.505.1053W} which includes multiple regularization algorithms for accurate few-body integration. {\tt\string Taichi} can be used along with any of the integration schemes present inside {\tt\string SpaceHub}. For this work, we found that the {\tt\string AR-Chain-Sym6+} regularization scheme is the most optimal. {\tt\string AR-Chain-Sym6+} is an updated {\tt\string AR-Chain} scheme which is more accurate than Mikkola's \citep{1999MNRAS.310..745M} original implementation. 

The improvements in {\tt\string AR-Chain-Sym6+} include an updated chain coordinate transformation, which improves on the CPU time taken to perform the coordinate transformation, active round-off error compensation, and the usage of a sixth order symplectic integration scheme instead of the traditional GBS extrapolation scheme used in the original {\tt\string AR-Chain} \citep{1999MNRAS.310..745M} method. This method is extremely efficient at handling highly eccentric systems. The usage of a fixed timestep maintains the time symmetry and as such helps achieve higher precision in round-off error dominated regime. For the same relative tolerance parameter, \cite{2021MNRAS.505.1053W} found that the {\tt\string AR-Chain-Sym6+} is at least 1-2 orders of magnitude better at conserving energy. For more information, we refer the interested reader to \cite{2021MNRAS.505.1053W}. {\tt\string Taichi} can be configured to allow an arbitrary number of particles to be treated by the regularization scheme. We performed tests and found that treating up to 20 particles with the regularization scheme was optimal in terms of performance and accuracy.

\subsection{Updates to FMM based Force Solver}

The multipole-to-local (M2L) kernel plays a crucial role in FMM by 
translating the multipole moments to local expansions for approximated force calculations. 
In the previous version of {\tt\string Taichi}, the order of 
expansions in the M2L kernel is kept at $p$, for both
multipole moments and the derivatives of $1/r$. After various optimizations, it is found that 
M2L kernel is evidently memory-bound instead of compute-bound. To increase
the efficiency of this kernel, we increase the expansion order for $1/r$ derivatives from $p$ to $2p$. This is called the \textit{double height} M2L 
kernel \citep[][]{Coulaud2008JCoPh.227.1836C} as opposed to the \textit{single height} formulation in our previous version. We found that this modifications improves the force error by a factor of $\sim 10\times$ for the same settings compared to the \textit{single height} version. 
As a result, we relax the force accuracy parameter $\epsilon$ by the same factor if double-height 
M2L is used. For more information, we refer the interested reader to \cite{Coulaud2008JCoPh.227.1836C}.

All of the improvements presented above enhance the accuracy and capability of {\tt\string Taichi}. We tested {\tt\string Taichi} with the initial conditions from \cite{2020MNRAS.493.3676O} and compared the results from {\tt\string NBODY6++GPU} \citep{2015MNRAS.450.4070W} to ensure correspondence between the two codes. The results are briefly presented in the Appendix. We found that {\tt\string Taichi} was able to simulate the systems as accurately as {\tt\string NBODY6++GPU}. For the purposes of our simulations, we found that an FMM input relative force accuracy parameter $\epsilon=2\times10^{-5}$ and a multipole expansion parameter $p=12$ was most optimal. For more information on these parameters, we refer the reader to \cite{2021ApJ...916....9M}. 

We ran tests with different values of $\epsilon$ and $p$ and found no difference in the final results. For our simulations we used a timestep parameter $\eta_{\rm T}=0.3$ unless more accuracy was demanded. For most simulations this results in a relative energy conservation of the order of $\sim$0.01\%. Under the presence of a dense segregated cusp, we found that the total relative energy conservation was $\sim$ 0.1\%-1\%. We note that no softening was used in the simulations. 
All of the simulations presented in this study were performed using only 32 threads on an {\tt\string AMD Epyc 7742} node. The simulations took $\sim 14-18$ days for $N \sim 1.32\times 10^6 $ to run to completion. The simulations with highly eccentric binaries or extremely dense cusps took much longer due to the formation of stable multiple systems.

\section{Models} \label{sec:models}
As mentioned in the previous section our choice of models is motivated by the work described in \cite{2020MNRAS.493.3676O}. We are interested in MBHs whose coalescence will be detectable by LISA and Tianqin. We set the mass of the primary $M_1=10^6 M_{\odot}$. The masses of the secondaries are generated such that we have mass-ratios $q=1.0,0.1,0.01$ where $q \equiv \frac{M_{2}}{M_{1}}$ and $M_{2}$ is the mass of the secondary. This enables us to systematically study the effect of the secondary on the collisonally relaxed cusp and vice-versa.

To generate the two-component models we assumed that the stars, white dwarfs, and neutron stars can be represented by a population of 1 $M_{\odot}$ particles, which are termed as the MS particles, while the heavier objects are represented using a population of 10 $M_{\odot}$ particles, which are called the BH particles. This is similar to the values used in contemporary studies \citep[][]{Preto2010ApJ...708L..42P,2012ApJ...744...74G}. To determine the fraction of MS particles to that of BH particles we consult \cite{2012ApJ...744...74G} and \cite{Antonini2014ApJ...794..106A}. Kroupa IMF \citep{Kroupa2001MNRAS.322..231K} predicts that
\begin{equation}
    N_{\rm MS} : N_{\rm BH} \sim 1:0.001
\end{equation}
However, we used values that are consistent with a top-heavy IMF \citep[e.g.,][]{Maness2007ApJ...669.1024M,2010ApJ...708..834B}. Spectroscopic data from late-type giants in inner parsec of Galactic Center provides evidence of continuous star formation consistent with that of a top-heavy IMF \citep{Maness2007ApJ...669.1024M}. In our simulations
\begin{equation} \label{equation:number_ratio}
    N_{\rm MS} : N_{\rm BH} = 1:0.005,
\end{equation}
similar to the value used in \cite{2012ApJ...744...74G}. Some IMFs predict an even higher fraction of BH particles \citep[e.g.,][]{Chabrier2005ASSL..327...41C}. The effect of changing the ratio of MS particles to BH particles would require further studies and is beyond the scope of this work. 

Each NSC in our simulation weighs $10^7 M_{\odot}$ in total, comprised of both MS and BH particles. Using the  IMF number ratio from equation \ref{equation:number_ratio}, we find that this implies that $M_{\rm BH} = 4.75 \times 10^5 M_{\odot}$ and $M_{\rm MS} = 9.525 \times 10^6 M_{\odot}$. In each NSC there are $N_{\rm MS}=655360$ and $N_{\rm BH}=3276$ particles. Thus, we model the system using a total of $N=1317272$ particles. Two additional particles are used to model the MBHs. To generate the $N$-body representations of the models, we use the self-consistent galaxy modeling toolkit {\tt\string Agama} \citep{Vasilev2019MNRAS.482.1525V}. In addition, {\tt\string Agama} includes a Fokker-Planck code called {\tt\string Phaseflow} \citep{Vasilev2017ApJ...848...10V} which we utilize to generate collisionally relaxed density profiles near MBHs. We describe the models in more details below. A brief summary of the initial conditions is provided in Tables \ref{tab:ic_summary} and \ref{tab:model_params}.  

\begin{table}
\centering
\begin{tabular}{|l|l|}
\hline
\textbf{Parameter}                         & \textbf{Value}                         \\ \hline
$N_{\rm MS}$      & $655360 $                       \\ \hline
$N_{\rm BH}$       & $3276 $                          \\ \hline
$N $                                & $1317272 $                       \\ \hline
$M_{\rm MS}$       & $9.525 \times 10^{6} M_{\odot}$ \\ \hline
$M_{\rm BH}$     & $4.75 \times 10^{5} M_{\odot}$  \\ \hline
$M_{\rm part;MS}$  & $14.5 M_{\odot}   $                       \\ \hline
$M_{\rm part;BH}$ & $145 M_{\odot}$                           \\ \hline

\end{tabular}%
\caption{Summary of the initial parameters used in the generation of the $N$-body models. }
\label{tab:ic_summary}
\end{table}

\subsection{Non-relaxed NSC Models}

To generate both the non-relaxed and the relaxed models, we start off with the Dehnen density profile \citep{Dehnen1993MNRAS.265..250D}  for both the MS and BH particles. The density profile for each component $i$ is given as:

\begin{equation}
    \rho_{i}(r) = \rho_0 \left( \frac{r}{r_{0}} \right)^{-\gamma_{i}} \left( 1+\frac{r}{r_{0}} \right)^{\gamma_{i}-4}
\end{equation}
where $\rho_0$ is the normalizing factor, $r_{0}$ is the scale radius and $\gamma_{i}$ determines the inner slope of the component $i$. We set $r_{0}=1.4$ pc following \cite{2020MNRAS.493.3676O}. We use $\gamma_{\rm MS} = \gamma_{\rm BH}=0.5$ which is the lowest value of $\gamma$ that can support an MBH at the center with an isotropic velocity distribution \citep{Baes2005A&A...432..411B}. We truncate the density profile at 1000 pc using an exponential truncation function. The distribution functions of both the MS and BH particles are generated by using the density function of each individual component and the combined potential of all components including the MBH, with an isotropic velocity distribution. This is all done under the self consistent framework of {\tt\string Agama}.

\subsection{Relaxed NSC Models}
\begin{figure}
    \centering
    \includegraphics[width=0.5\textwidth]{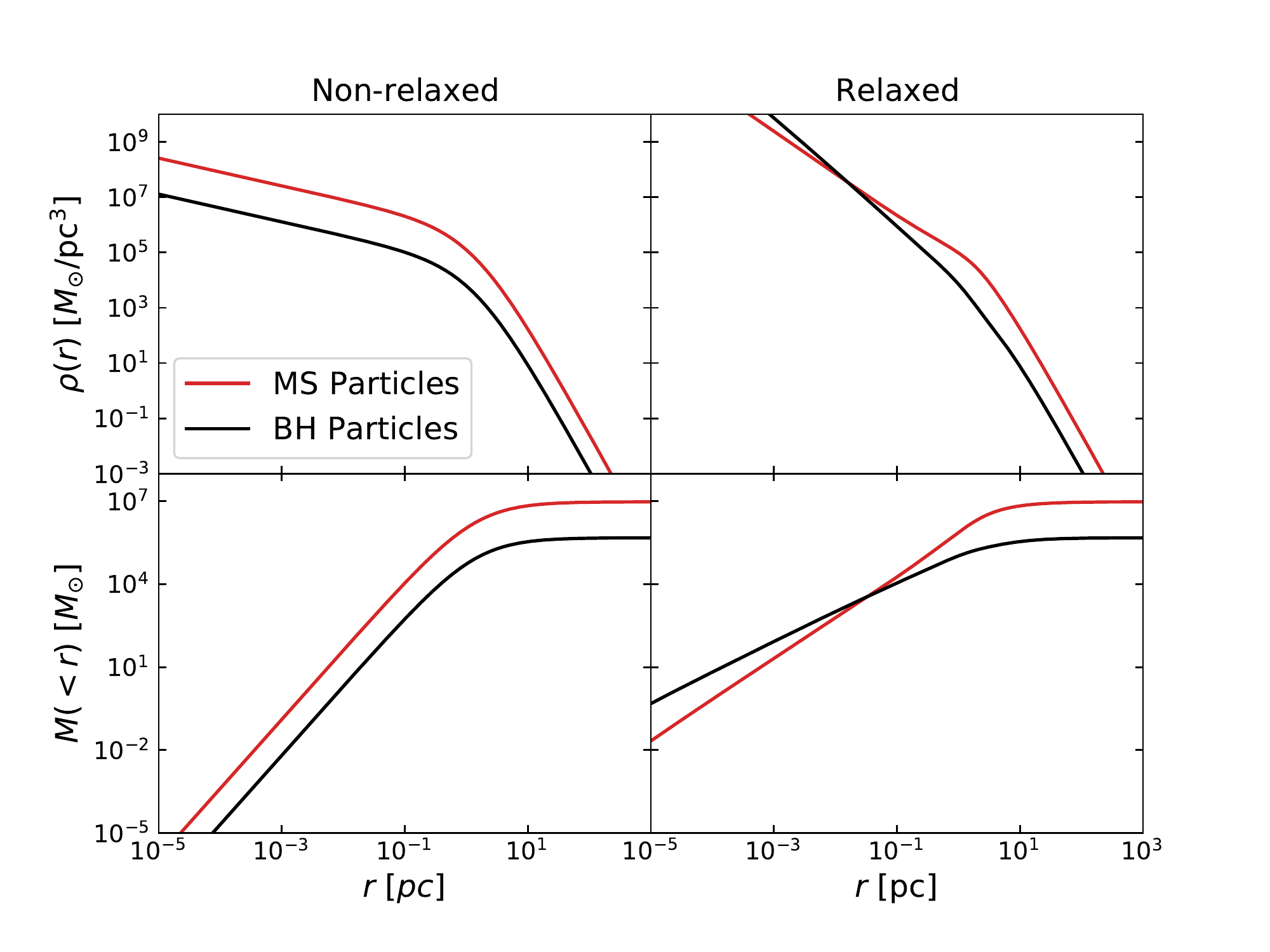}
    \caption{The analytic density $\rho(r)$ and the enclosed cumulative mass $M(<r)$ a function of $r$, the distance from the center of the cluster under the presence of a $10^6 M_{\odot}$ MBH at the center. The analytic profiles have been computed using {\tt\string Phaseflow} The differences in the relaxed and the non-relaxed cases are evident with collisional relaxation implying mass segregation. The relaxation produces a denser cusp near the $10^{6} M_{\odot}$ MBH and stellar mass black holes dominate the total mass for all radii $< 0.1$ pc. The MBH is dominant in regions with $r<1$ pc.}
    \label{fig:init_conditions}
\end{figure}

To generate the collisionally relaxed models, we have to follow a few more steps. We input the non-relaxed density profiles of both the MS and BH particles along with the mass of the MBH at the center. Then, we evolve the system using the Fokker-Planck code {\tt\string Phaseflow} until the system has evolved to a collisionally relaxed state. For the model with a $10^6 M_{\odot}$ MBH at the center, the relaxed state is achieved in $\sim 0.5$ Gyr.  This occurs when the inner density profile of the BH particles falls off as $\sim r^{-2}$ and that of the MS particles falls off as $\sim r^{-1.5}$ \citep[][]{1977ApJ...216..883B, Hopman2006ApJ...645L.133H,2009ApJ...697.1861A}. The output from {\tt\string Phaseflow} can be easily used to generate isotropic models using {\tt\string Agama} in a fashion similar to the one described above. We present a comparison of the analytic density and mass profiles in Figure \ref{fig:init_conditions} for the non-relaxed and the relaxed cases when we have a $10^6 M_{\odot}$ SMBH at the center. In both the non-relaxed and the relaxed cases, we verified the $N$-body models accurately reproduced the analytic density and mass profiles.  

\subsection{Generating the Merger Models}

To initialize the merger between two NSCs, we follow the steps outlined in \cite{2020MNRAS.493.3676O}. The two NSCs and their corresponding MBHs are initially unbound and are allowed to become bound over the course of the simulation. The initial separation between the two MBHs is denoted as $d_{\rm in}$. In our simulations we set $d_{\rm in} = 20$ pc. We verified that this is less than the effective radius of the NSCs. To generate the initial relative velocity of the two NSCs, we use a free parameter $\xi$ similar to the parameter $\eta$ described in \cite{2020MNRAS.493.3676O} equation (9). $\xi$ quantifies the initial angular momentum of the orbit. Smaller values of $\xi$ imply a more eccentric orbit. In our models, the relative velocity $v_{\rm in}$ is defined as
\begin{equation}
    v_{\rm in} = \xi \sqrt{\frac{GM_{*}(d_{\rm in})}{d_{\rm in}}}
\end{equation}
where $M_{*}(d_{\rm in}) $ accounts for the total mass (excluding the MBH masses) within a distance of $d_{\rm in}$ from the center of each NSC. Once $v_{\rm in}$ has been obtained, the NSC of the secondary along with its MBH is placed at a position centered around $\mathbf{r}_{\rm in} =  \left(d_{\rm in},0,0\right)$ with a velocity $\mathbf{v}_{\rm in} = \left(0,v_{\rm in},0\right)$, while the other NSC is place at the origin with zero bulk velocity. In each simulation, we verified that the initial relative velocity was the same to maintain consistency. We found that using $d_{\rm in} = 20$ pc, $v_{\rm in} \approx 55.5$ km/s. Six simulations are generated with the relaxed and the non-relaxed NSCs with $\xi=1.0$.  We label the simulations where relaxed NSCs are used as {\tt\string r\_} simulations and the simulations where non-relaxed NSCs are used as {\tt\string nr\_}. To reduce the number of simulations performed due to computational constraints we perform mergers of relaxed NSCs with only other relaxed NSCs and non-relaxed NSCs with other non-relaxed NSCs. We expect the results of mixed simulations to lie in-between the results obtained in this study. 

We expect most mergers to happen on eccentric orbits rather than circular. Thus, it is imperative to understand the effects of initial eccentricity and its evolution under the presence of relaxed and non-relaxed cusps. To understand the effects of an initially eccentric orbit, we perform four additional simulations with $q=0.1$, $\xi=0.5$ and $q=0.1$, $\xi=0.1$. The former simulations are labelled {\tt\string ecc\_1} whereas the latter simulations are labelled {\tt\string ecc\_2} 
The circular orbit models are used to understand how the evolution of the binary changes as a function of the mass-ratio while the eccentric models are used to understand how the eccentricity is affected by the different density profiles for a given mass ratio.

To demarcate the three stages of evolution, we need the influence radius and the hard binary radius. In order to determine the influence radius and the hardening radius, we consult \cite{2013degn.book.....M} equation (8.71). 
 We use the following definitions which are more suitable for $N$-body simulations; once a bound binary is formed, an influence radius of the primary can be defined as  
\begin{equation} \label{equation:influence_rad_binary}
    d_{\rm infl} \equiv r_{\rm enc}( 2 M_{1})
\end{equation}
where $r_{\rm enc}$ is the radius enclosing the amount of mass in the parenthesis.
 The corresponding hard binary radius can be defined then as
\begin{equation} \label{equation:hard_binary_binary}
    a_{\rm h} = \frac{q}{(1+q)^2} \frac{d_{\rm infl}}{4}.
\end{equation}
For more definitions, we refer the reader to \cite{2006ApJ...648..890M}.

All the simulations were run for a total time of 10 Myr. In most simulations, this was enough for the MBH binary to harden to $a_{\rm h}/5$ which is usually sufficient to study the effects of core scouring as reported in previous studies \citep[e.g.,][]{2006ApJ...648..890M}.

\begin{table}
\centering
\begin{tabular}{|l|l|l|l|l|}
\hline
\textbf{Simulation ID}   & $\gamma_{\rm MS}$ & $\gamma_{\rm BH}$ & $q$    & $\xi$ \\ \hline
{\tt\string r\_q\_1.0}        & 1.5       & 2.0       & 1.0  & 1.0                 \\ \hline
{\tt\string r\_q\_0.1}        & 1.5       & 2.0       & 0.1  & 1.0                 \\ \hline
{\tt\string r\_q\_0.01}        & 1.5       & 2.0       & 0.01 & 1.0                 \\ \hline
{\tt\string nr\_q\_1.0}      & 0.5       & 0.5       & 1.0  & 1.0                 \\ \hline
{\tt\string nr\_q\_0.1 }     & 0.5       & 0.5       & 0.1  & 1.0                 \\ \hline
{\tt\string nr\_q\_0.01}      & 0.5       & 0.5       & 0.01 & 1.0                 \\ \hline
{\tt\string r\_q\_0.1\_ecc\_1}   & 1.5       & 2.0       & 0.1  & 0.5                 \\ \hline
{\tt\string r\_q\_0.1\_ecc\_2} & 1.5       & 2.0       & 0.1  & 0.1                 \\ \hline
{\tt\string nr\_q\_0.1\_ecc\_1}   & 0.5       & 0.5       & 0.1  & 0.5                 \\ \hline
{\tt\string nr\_q\_0.1\_ecc\_2} & 0.5       & 0.5       & 0.1  & 0.1                 \\ \hline
\end{tabular}%
\caption{Summary of the model parameters used for the NSC-NSC merger simulations. All of the above simulations use the same number of particles. The first six models are on  circular orbits initially and used to study the effect of relaxation as a function of $q$ and the last four models are eccentric and used to study the effect of initial eccentricity and evolution of eccentricity at a fixed $q$. {\tt\string ecc\_1} models are moderately eccentric whereas {\tt\string ecc\_2} models are highly eccentric.}
\label{tab:model_params}
\end{table}

\section{Results} \label{sec:results}
\begin{figure*}
\centering
\includegraphics[width=0.8\textwidth]{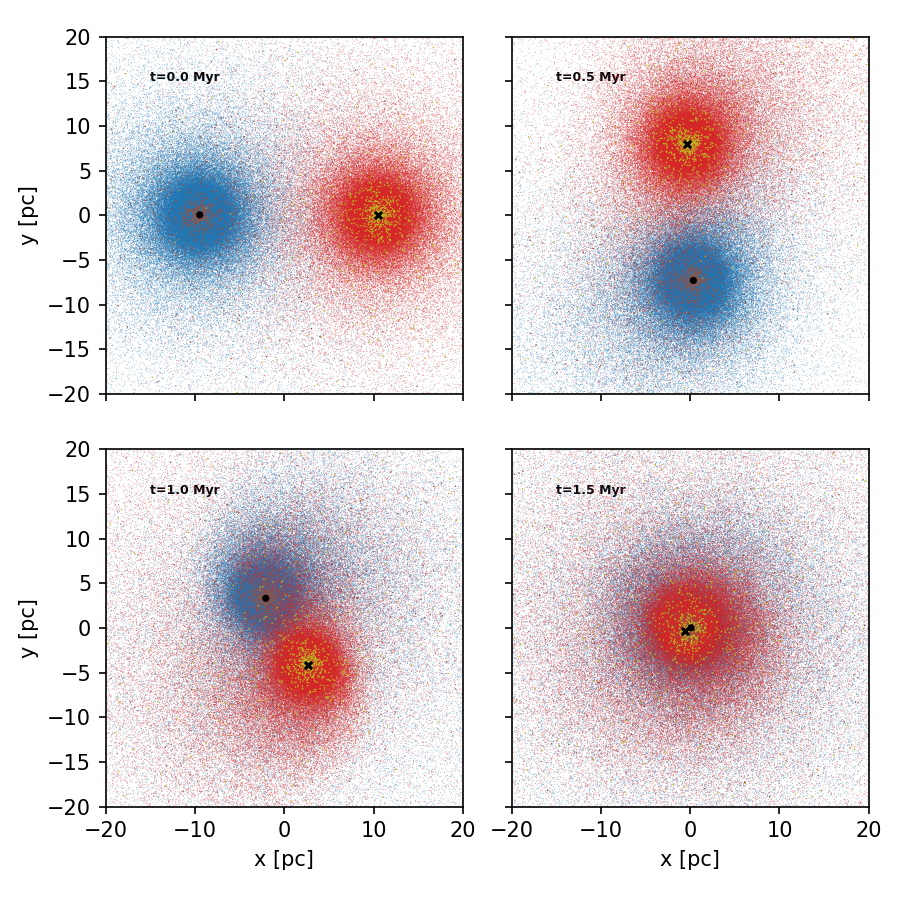}
\caption{A scatter plot of the two NSCs with MBHs projected onto the $x-y$ plane at different points in time during the merger process. The simulation being pictured here is {\tt\string r\_q\_0.1}.  
As the simulation proceeds, the NSCs belonging to the primary (black circle) and secondary (black cross) are brought closer to each other by the combined effects of dynamical friction and tidal forces from stripped stars leading to a mixture of the MS (blue, red dots) and BH particles (brown, yellow dots) from both NSCs. 
The NSCs merge within $\sim 1.5$ Myr resulting in the formation of a hard binary at the center.}
\label{fig:part_scatter_plot}
\end{figure*}

\begin{figure*}
\includegraphics[width=1.0\textwidth]{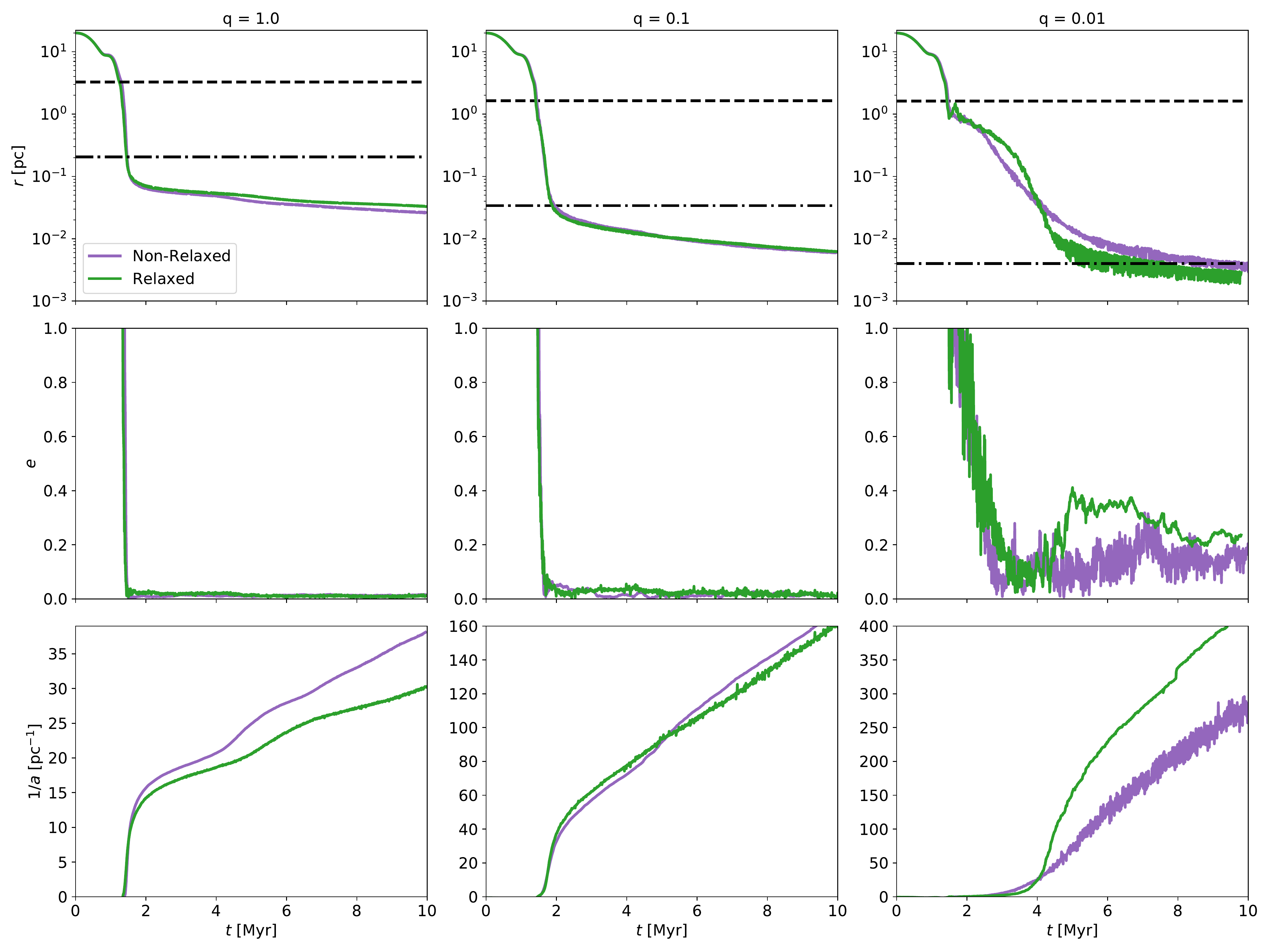}
\caption{The evolution of the binary parameters as a function of time for the  circular orbit models. The dashed line represents the influence radius of the binary and the dash-dotted line represents the hard-binary radius. Top: evolution of the separation $(r)$ between the two MBHs as a function of time. Middle: evolution of the eccentricity ($e$) as a function of time. Bottom: evolution of the inverse semi-major axis $(1/a)$ as a function of time. The different evolutionary tracks between the non-relaxed and the relaxed cases highlight the imprint of the surrounding NSC on the dynamics of the MBH binary. We find that while non-relaxed models reach hard binary radius and harden faster for $q=1.0$, the opposite happens for $q=0.01$.}
\label{fig:merger_profile_all}
\end{figure*}

To visually examine the evolution of the MBH binary over time, we present snapshots of the evolution over the first $1.5$ Myr for the {\tt\string r\_q\_0.1} model in Figure \ref{fig:part_scatter_plot}. The orbit of the MBHs is in the $x-y$ plane, and the plot's origin is the center of mass. We find that the NSCs merge within $\sim 1.5$ Myr leading to the formation of a hard MBH binary. To quantitatively understand the dynamics in more detail, we plot the evolution of the orbital separation, eccentricity, and inverse semi-major axis of the binary over time in Figures \ref{fig:merger_profile_all} and \ref{fig:q_01_ecc}. They helps us understand the differences between relaxed and non-relaxed models as a function of the mass ratio $q$ and the initial eccentricity. We analyze each step of the evolution in the sections below. We first present our analysis of the circular orbit models before moving on to the eccentric ones.

\subsection{Pre-binary phase ($r > d_{\rm infl}$)}
\label{subsec:res_prebinary}
The first stage of evolution, the pre-binary phase, lasts until a bound binary has formed. 
Examining the orbital elements in Figure \ref{fig:merger_profile_all}, we find that this phase lasts for the first $\sim 1.5$ Myr for both relaxed and non-relaxed models across different $q$. The NSCs merge in roughly $\sim 1.5$ Myr bringing the secondary to within the influence radius of the primary ($\sim 1$ pc) leading to the formation of a bound binary.

In this phase the evolution is dominated by the dynamical friction of the stars and the drag force from the tidally stripped stars which help in reducing the angular momentum of the MBHs \citep{2020MNRAS.493.3676O}. The latter effect is more important in this stage as dynamical friction typically acts on longer timescales.  The process involves the transfer of angular momentum from the NSC cores to the stripped stars which expand their orbits. As demonstrated by \cite{Huang1963ApJ...138..471H} and later by \cite{2020MNRAS.493.3676O}, if the two NSCs are considered to be part of a binary system, then in the event of a mass loss, the change in specific angular momentum $l$ can be written as 
\begin{equation}
    \delta l = \left(l_{s} - l\right)\frac{\delta m_{s}}{m}
\end{equation}
where $m$ is the mass of the NSC binary system and $m_s$ and $l_s$ are the mass and specific angular momentum of the stars that have been tidally stripped from the NSC. $\delta m_s$ is large in the early phase of the dynamical evolution, while it can be negligible in the later phase. Under the assumption that the eccentricity of the stripped stars has not changed and that the mass loss through tidal disruption is negligible compared to that of the mass of the NSC binary, we can write the expressions for the specific angular momentum of the NSC binary and that of the stripped stars as follows:

\begin{gather}
    l = \sqrt{Gma(1-e^2)}\\
    l_s = \sqrt{Gm(a+\delta a)(1-e^2)}.
\end{gather}
In order to satisfy the condition that $\delta m_s < 0$ and $\delta l < 0$, we find that $\delta a > 0$.
Thus, as the expansion of orbit is associated with an increase in angular momentum, and the total angular momentum remains conserved, the distances between the NSC cores, and MBHs embedded in them, shrink as a result. 
This mechanism is especially important for lower $q$ cases as dynamical friction works inefficiently to decay the orbit of less massive BHs \citep{2020MNRAS.493.3676O}. 
The results are consistent with those presented in \cite{2020MNRAS.493.3676O}. This is not surprising since we compared the amount of mass losing and gaining angular momentum between simulations and found that they were equivalent. This leads to negligible differences in the evolution during this period.
However, we warn the readers that the effect of tidal stripping would be reduced in realistic galaxies. Interestingly, we find that the time of binary formation, which is dictated by the tidal effects, is consistent amongst simulations performed with $N = 4\times10^{5}$ and $N=1.32 \mathrm{m}$ (see appendix \ref{appendix:resolution}) suggesting that a lower resolution is sufficient enough to resolve tidal stripping effects and time of binary formation. A systematic study of the effect of resolution on tidal stripping is beyond the scope of this work.

\subsection{Bound binary phase ($d_{\rm infl} < r < a_{\rm h} $)} \label{subsec:combined_phase}

In the second phase of evolution, after a bound binary has formed, orbital decay occurs due to a mix of dynamical friction and three-body scattering events. When the two MBHs are sufficiently far apart, dynamical friction acts on each body independently to shrink the binary \citep[][section 8.2.2]{2013degn.book.....M}. When the binary gets closer, hardening via scattering becomes more important and the efficiency of scattering depends on the binary mass ratio $q$ \citep[e.g.,][section 8.2.2]{2013degn.book.....M}. 

Examining Figure \ref{fig:merger_profile_all} we find that for $q=1.0,0.1$, the combined phase proceeds extremely quickly leading to the formation of the hard binary immediately after the end of the first phase. The evolution in the $q=0.01$ models is more gradual. In addition, we find some notable differences in this phase between the relaxed and the non-relaxed models that depend on  $q$. We notice from the evolution of the inverse semi-major axis that in {\tt\string nr\_q\_1.0}, the binary is able to settle at a smaller separation than in {\tt\string r\_q\_1.0}. This is quite surprising since intuitively we would expect the denser relaxed cusp to yield a faster orbital decay. As the mass ratio is lowered, the situation changes and in the $q=0.01$ case, we find that the the relaxed model actually accelerates the transition to the hard binary stage. Since the orbital energy of the binary is given as 
\begin{equation}
    E_{\rm binary} = -\frac{GM_{\rm 1}M_{\rm 2}}{2a}
\end{equation}
where $a$ is the semi-major axis of the binary, we deduce that for for the $q=1.0$ case, the binary is able to lose more energy in the non-relaxed model compared to that in the relaxed model. As we lower the mass ratio, the energy loss in the relaxed models increases. We seek to understand the processes in work that change the results across mass-ratios.

In order to understand the differences, we first approximately quantify the amount of energy lost by dynamical friction and scattering. We follow \cite{2013degn.book.....M} equations (8.73) and (8.74) which state
\begin{equation} \label{equation:dedtdf}
    \frac{dE}{dt} \bigg|_{\rm df} \approx -4.4 \frac{G^2 M_{2}^{2} \rho(r) \rm{ln}(\Lambda)}{\sigma}
\end{equation}
and 
\begin{equation} \label{equation:dedts}
    \frac{dE}{dt} \bigg|_{\rm s} = -\frac{H(a)}{2q} \frac{G^2 M_{2}^{2} \rho(r)}{\sigma} 
\end{equation}
where $\frac{dE}{dt} \big|_{\rm df} $ and $\frac{dE}{dt} \big|_{\rm s}$ are the energy losses from dynamical friction and scattering respectively, $\rm{ln}(\Lambda)$ is the Coulomb logarithm, $\sigma$ is the velocity dispersion and $H(a)$ is the dimensionless scattering rate. As the binary hardens, the energy losses via scattering become more important as the scattering efficiency increases as $q^{-1}$. Physically, this makes sense since the larger the mass of the secondary MBH, the more energy would have to be extracted by the intruder to harden the binary by a fixed amount.

From equations \ref{equation:dedtdf} and \ref{equation:dedts} we notice that $\frac{dE}{dt} \propto M_{2}^{2}$. Thus, the rate of loss of energy is faster for binaries with larger secondary mass, which explains why the combined phase proceeds rapidly for $q=1.0,0.1$ in contrast to the $q=0.01$ case. Dynamical friction is less efficient in the $q=0.01$ models leading to a more gradual orbital decay. However, this does not explain why the orbital decay is more efficient in the {\tt\string nr\_q\_1.0} model compared to the {\tt\string r\_q\_1.0} model.

\begin{figure}
    \centering
    \includegraphics[width=0.4\textwidth]{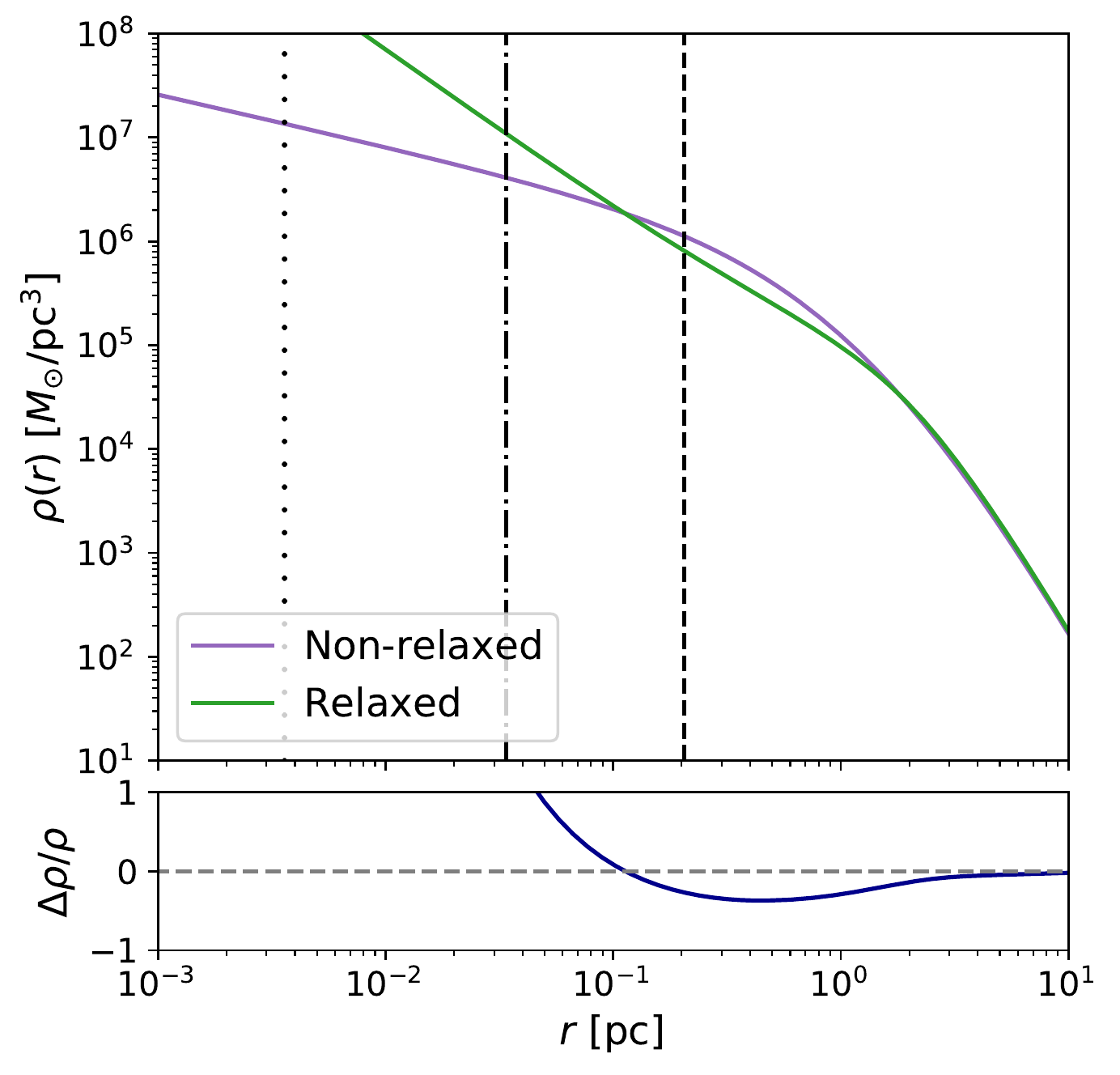}
    \caption{The initial density $\rho(r)$ and relative density difference $\Delta \rho / \rho$ of MS particles as a function of $r$, the distance from the center of the cluster under the presence of a $10^6 M_{\odot}$ MBH at the center. The dashed line, dash-dotted line, and the dotted line represent the hard-binary radii for the $q=1.0,0.1,0.01$ models respectively. We find that for all radii within the influence radius of the primary, the density of the MS particles is lower in the relaxed models compared to the non-relaxed models until we reach $\sim 0.1 d_{\rm infl}$}
    \label{fig:ms_ic_compare}
\end{figure}

To understand the differences in evolution of the inverse semi-major axis between the non-relaxed and relaxed models we note that $\frac{dE}{dt} \propto \rho(r)$. Therefore, we focus on the differences in the initial density profile of the non-relaxed and the relaxed models. For clarity, we compare the density profile for the MS particles in the case where the central MBH mass is $10^{6} M_{\odot}$ in Figure \ref{fig:ms_ic_compare}. Since the total mass of the MS particles is much more than that of BH particles, we expect any major discrepancies to arise out of differences in the MS density and mass profiles. 

Looking at all radii $<10$ pc, we find that in the relaxed models, the density of MS particles is lower than that of the non-relaxed models for all radii $>0.1$ pc. Since the NSC mass across our models is fixed and the relaxed models have higher central density, the density in the outskirts decreases. This counter-intuitive result was also reported in \cite{2012ApJ...744...74G}. To explain it, the authors accounted for the effect of the BH particles on the MS particles near the SMBH. Since the BH particles are more massive than the MS particles, they dynamically heat the MS particles leading to a lower density in the above-mentioned region. However, \cite{2012ApJ...744...74G} showed that at smaller radii, due to the scattering effects of the BH particles, the MS particles end up forming a denser, Bahcall-Wolf cusp. This cusp has a higher density only for very small radii, typically $\sim 0.1d_{\rm infl}$. 

As the MBH binary hardens to $a_{\rm h}$ and energy loss due to scattering becomes more dominant, differences in evolution appear between the non-relaxed and the relaxed models.  We find that $a_{\rm h} = 0.205 $ pc for $q=1.0$, the largest amongst all our models. This lies in the region where the density $\rho(r)$ for the relaxed models is lower than that of the non-relaxed models implying that the energy loss in the non-relaxed models must be higher for {\tt\string nr\_q\_1.0}.  

The situation changes as we lower the mass-ratio and the hard-binary radius decreases. We find that the total density $\rho$ increases after $r \sim 0.1$ pc. Owing to the higher density, the binary is able to compensate or even more than compensate for the differences in the energy loss in the beginning of the combined stage in the $q=0.1,0.01$ models. This is quite evident while comparing the evolution of the binary separation for non-relaxed and relaxed $q=0.01$ models. The initial inspiral of the secondary is slower in {\tt\string r\_q\_0.01} compared to {\tt\string nr\_q\_0.01}. Once the secondary is about $\sim 0.1$ pc away, the inspiral accelerates and it reaches the hard binary radius faster than the non-relaxed model. The situation is quite similar to the {\tt\string RUN3} model in \cite{Khan2015ApJ...798..103K} where the authors found that initial inspiral of the secondary for the more centrally concentrated model to be slower in the beginning. In the semi-analytic model developed in \cite{Gualandris2022MNRAS.511.4753G}, the authors found that the inspiral due to dynamical friction was slower in the models with steeper inner cusps due to lower stellar density in the outskirts.

Incidentally, for the chosen set of initial conditions, {\tt\string r\_q\_0.1} and {\tt\string nr\_q\_0.1} show almost identical evolution of binary parameters. For $q \leq 10^{-1}$, higher central concentration drives the binary towards faster inspiral. For really low mass ratio binaries this has the potential to accelerate transition to hard-binary stages even faster but further studies with higher resolution models are required.

\begin{figure*}
\centering
\begin{subfigure}[b]{\linewidth}
\includegraphics[width=\linewidth]{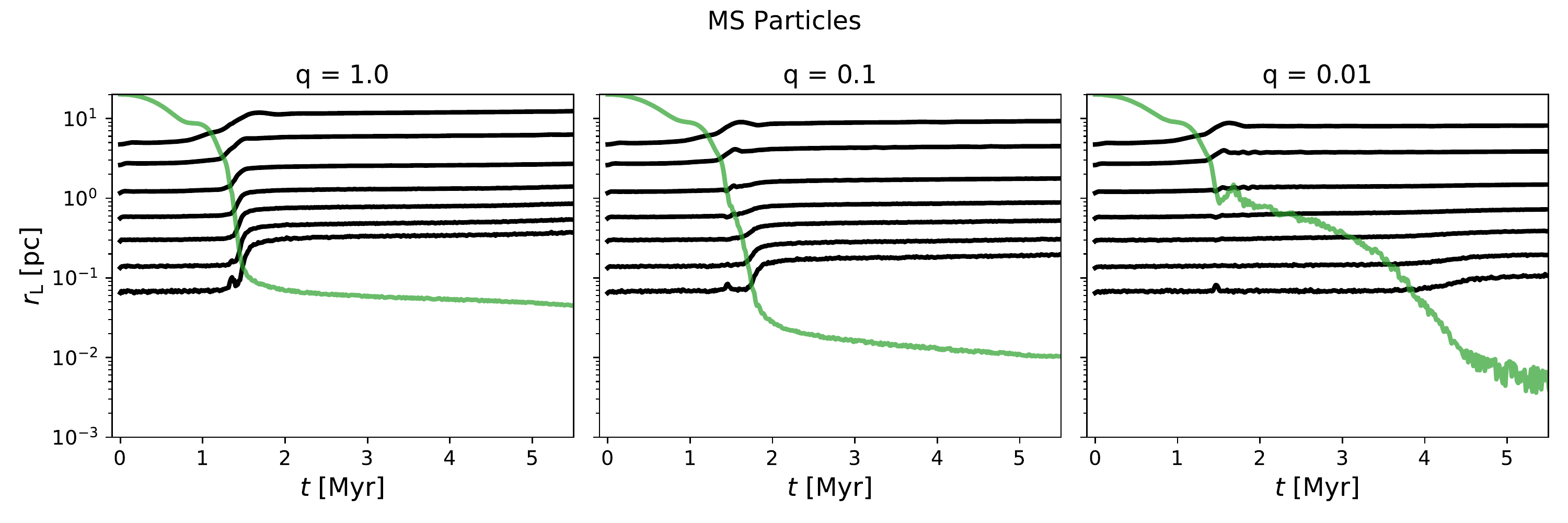}
\end{subfigure}

\begin{subfigure}[b]{\linewidth}
\includegraphics[width=\linewidth]{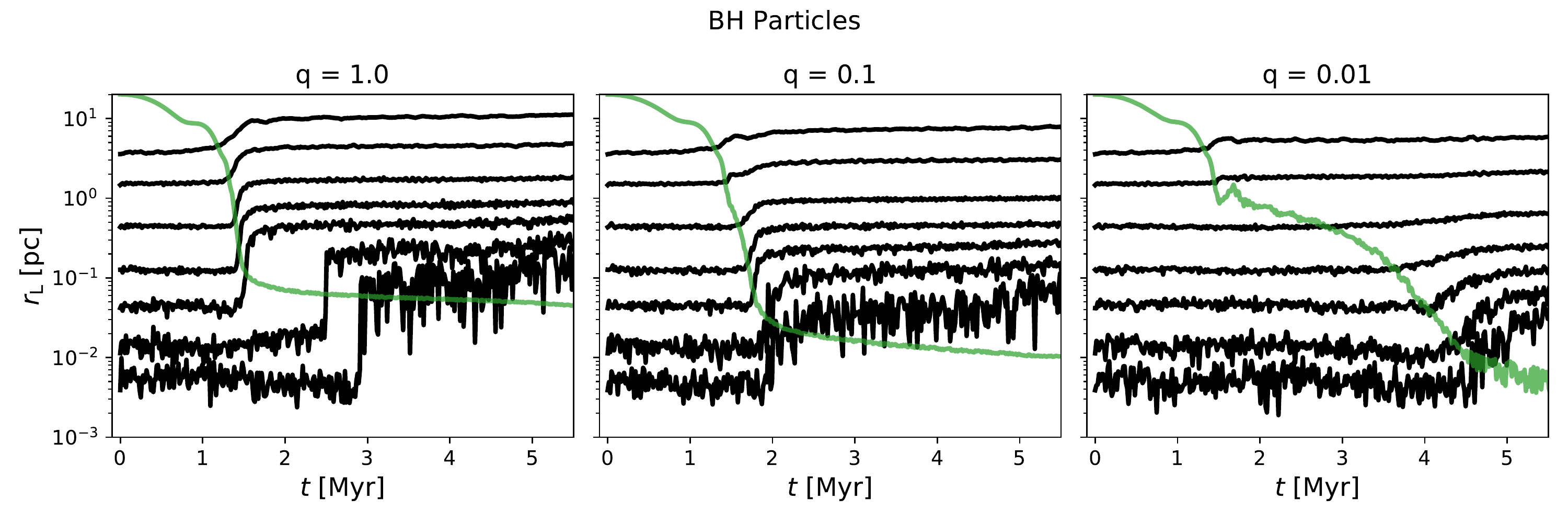}
\end{subfigure}

\caption{The evolution of the Lagrange radius $(r_{\rm L})$ of different mass-fractions as a function of time $(t)$ for the relaxed models. The green curves denote the separation of the MBHs as a function of time. For clarity, only particles belonging to NSC 1 have been taken into account here. From bottom to top the mass fractions are 0.1\%, 0.3\%, 1\%, 3\%, 10\%, 30\%, and 50\%. Top: The evolution of the Lagrange radius for the MS particles. Bottom: The evolution of the Lagrange radius for the BH particles. We notice that as the mass ratio decreases, the cusp is perturbed less, and mass-segregation is only partially reversed. The results are qualitatively consistent with similar ones presented in \citet{2012ApJ...744...74G}. }
\label{fig:lagrangian_main_seg}
\end{figure*}

As a back-reaction to the shrinkage of the binary due to dynamical friction and scattering, energy is induced into the particles nearby leading to an expansion of their orbits and therefore, a disruption of the cusp. Figure \ref{fig:lagrangian_main_seg} provides a visual description of said expansion. Comparing the Lagrange radii, i.e. the radii enclosing a particular fraction of the total mass, plots of both MS and BH particles, we find that in the combined phase, higher mass ratios inject more energy into the surrounding particles leading to a rapid expansion and disruption of the cusp, which also leads to reversal of mass-segregation. The disruption is less-violent in $q=0.01$ case because of the lower mass of the secondary.

Since the MBHs form a bound binary, we can also examine the evolution of eccentricity during this stage. For the larger mass-ratio models we find that the binary's orbit is approximately circular as it reaches the hard binary radius. We see similar trends in both relaxed and non-relaxed models for $q=1.0,0.1$. This is similar to the observation made by \cite{2012ApJ...744...74G} in their circular orbit models.  For $q=0.01$, the story is a little different. While the binaries initially start circular, the growth of eccentricity is more stochastic in this case because of the lower mass of the secondary. However, we find that evolution of eccentricity is generally in the direction of higher eccentricity. This was also noted in \cite{2013degn.book.....M} for Intermediate Mass Black Holes (IMBHs) in this mass range. Curiously, we find that for {\tt\string r\_q\_0.01}, the binary is able to reach a higher eccentricity in this phase compared to {\tt\string nr\_q\_0.01}. Intuitively, we would expect the opposite because because when the periapsis of the binary falls within the denser, relaxed cusp, it should circularize the binary. The increase in eccentricity happens around the time the secondary reaches the distances where the mass is dominated by BH rather than MS particles. The secondary hardens by having more encounters with the BH particles. In asymmetric MBH binaries the eccentricity growth is mostly driven by the companion-perturber mass ratio \cite{Sesana2008ApJ...686..432S}. A larger intruder mass in the relaxed model results in growth of eccentricity. This may not be physical because in realistic NSCs, the mass of the perturbing BH particles would be smaller. 

We also noted that the evolution of eccentricity for $q=0.01$ model is quite dependent on resolution. We found that a simulation using $N\sim 4\times 10^{5}$ particles (see appendix \ref{appendix:resolution}) resulted in the MBH binary reaching extremely large values of eccentricity (0.9) in both non-relaxed and relaxed models. It highlights the importance of using larger resolution to estimate binary evolution parameters, especially for lower mass ratio binaries. Our results for this mass range are quite similar to those presented in \cite{ArcaSedda2018MNRAS.477.4423A} where the authors found that unless the IMBH starts off in an highly eccentric orbit, it is not able to reach large values of eccentricity. We caution against consulting single simulations to track the evolution of eccentricity with the current resolution. Unlike semi-major axis, the evolution of eccentricity is a second order effect in angular momentum and subject to more stochasticity and more simulations are needed to model the evolution more realistically. 

\subsection{Hard-binary phase ($ r < a_{\rm h} $)}

The last phase before the GW emission state is the hard-binary phase where the binary hardens by three-body scattering. 
We investigate the differences in hardening rates between the relaxed and non-relaxed models. In the full loss-cone regime the binary should harden at a fixed rate. This would imply that,
\begin{equation}
    \frac{d}{dt} \left(\frac{1}{a}\right) = s
\end{equation}
where $s$ is some constant. To find the value of $s$, we fit straight lines to the inverse semi-major axis plots. For $q=1.0,0.1$ we fit the lines to the values between 5 Myr and 10 Myr. For $q=0.01$, we do it between 7 Myr and 10 Myr since the binary reaches the hard binary radius a little before 7 Myr. We notice that there is a sharp jump in the evolution of the inverse semi-major axis in the {\tt\string r\_q\_0.01} model around $\sim 8$ Myr. Jumps in the evolution of the inverse semi-major axis was also noted in \cite{Khan2018A&A...615A..71K}. The authors attributed them to the ejection of an intruder following the formation of a triple system between the MBH binary and a third heavy particle. In our instance, this third particle happened to be a BH particle. The jumps do not dominate the overall evolution. To ensure consistency, the average value of $s$ in the non-jump regions is taken to be the hardening rate in this particular scenario. 

The values of $s$ found for the different simulations are reported in Table \ref{tab:slope_summary}. For the $q=1.0$ models, the hardening rate in non-relaxed scenario is  $\sim 35$\% higher than that in the relaxed scenario whereas in the $q=0.1$ models the hardening rates of both relaxed and non-relaxed models are within $\sim 10$\% of one another. In the $q=0.01$ models, the hardening rate of the non-relaxed model is $22\%$ lower than that in the relaxed model.
We notice that the value of $s$ does not change much between relaxed and non-relaxed models indicating that even though the cusp might be disrupted less in lower mass cases, the primary mode of evolution in this phase is collisionless and dependant on the geometry of the merger product. The lower hardening rate in {\tt\string r\_q\_1.0} is quite interesting. To understand it, we looked at the density profile during the hardening phase and found that the density profile in {\tt\string r\_q\_1.0} was $\sim$20\% lower than that in {\tt\string nr\_q\_1.0} at the influence radius ($\sim 1$ pc) of the binary. Since this is where the stars in the loss cone arise out of, the lower scattering rate is caused to to fewer particles in the loss cone in the {\tt\string r\_q\_1.0} model.

\begin{table}
\centering
\begin{tabular}{|l|l|}
\hline
\textbf{Simulation ID} & $s $ $[\rm{pc}^{-1} \rm{Myr}^{-1}]$\\ \hline
{\tt\string r\_q\_1.0}        & 1.90      \\ \hline
{\tt\string r\_q\_0.1}        & 13.9     \\ \hline
{\tt\string r\_q\_0.01}        & 46.6     \\ \hline
{\tt\string nr\_q\_1.0}      & 2.6      \\ \hline
{\tt\string nr\_q\_0.1}      & 14.6     \\ \hline
{\tt\string nr\_q\_0.01}     & 39.4     \\ \hline
{\tt\string r\_q\_0.1\_ecc\_1}   & 8.2     \\ \hline
{\tt\string nr\_q\_0.1\_ecc\_1} & 9.2     \\ \hline
{\tt\string r\_q\_0.1\_ecc\_2} & 16.2     \\ \hline
{\tt\string nr\_q\_0.1\_ecc\_2} & 15.7     \\ \hline

\end{tabular}%
\caption{Summary of the slopes of the inverse semi-major axes of the various simulations. We find that hardening rates between relaxed and non-relaxed models are within $\sim$30\% of each other contrary to the findings of \citet{Khan2018A&A...615A..71K}. In addition, we find that {\tt\string nr\_} models harden faster for $q \geq 0.1$ whereas the opposite is observed for $q=0.01$}
\label{tab:slope_summary}
\end{table}

We also examine the growth of eccentricity during this phase of evolution. $q=1.0,0.1$ models remain roughly circular whereas the $q=0.01$ models show stochasticity in the evolution of eccentricity. We find that {\tt\string nr\_q\_0.01} shows some mild growth resulting in a final eccentricity of 0.2 by 10 Myr whereas the eccentricity actually decreases in {\tt\string r\_q\_0.01}. This is quite curious and we leave the investigation to a future study.

\subsection{Eccentric Orbital Parameters}

\begin{figure*}
    \centering
    \includegraphics[width=1.0\textwidth]{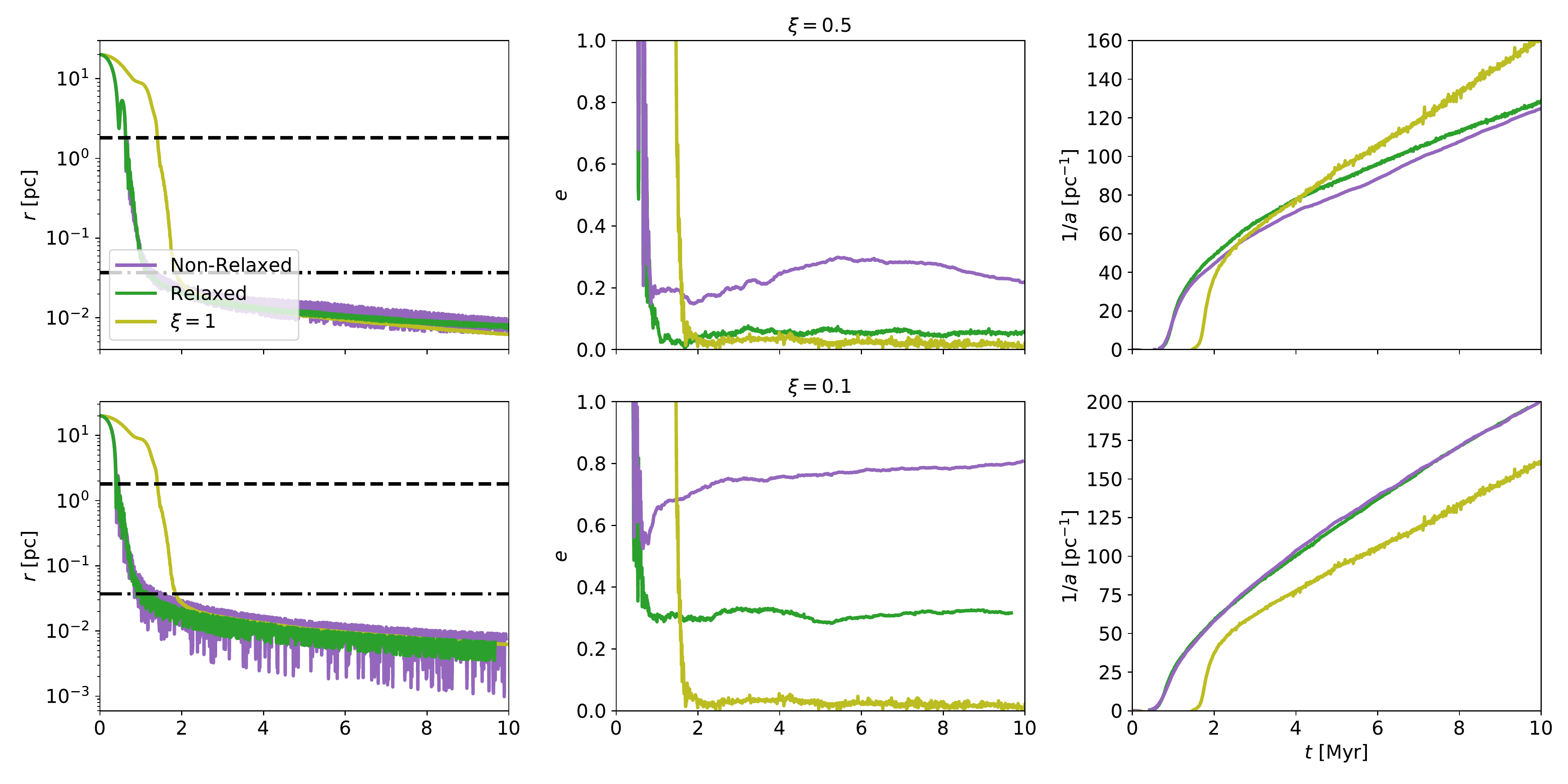}
    \caption{ The evolution of the binary orbital parameters for the eccentric models as a function of time $t$ in Myr. Top: {\tt\string ecc\_1} models that are mildly eccentric initially. Bottom: {\tt\string ecc\_2} models that are highly eccentric initally. We find similar trends as we found for the  circular orbit models of $q=0.1$ in Figure \ref{fig:merger_profile_all} for the binary hardening rates. We, however, notice that the presence of the denser relaxed cusp affects the eccentricity evolution of the binary. The relaxed cusp circularizes the binary more than the non-relaxed cusp. This is more evident in the highly eccentric scenario (bottom) where the binary in the relaxed cusp forms at much lower eccentricity and does not show any growth over time. }
    \label{fig:q_01_ecc}
\end{figure*}

Realistic galaxy mergers are more likely to happen on eccentric orbits. The aforementioned parameter $\xi$ can be changed to change the initial amount of angular momentum. Reducing the parameter can lead to more eccentric orbits. We studied the effect of eccentric initial conditions and how they interplay with relaxed and non-relaxed models. Figure \ref{fig:q_01_ecc} shows that the evolution of the separation and the binary hardening is qualitatively identical to the initial conditions of $\xi = 1$ and $q=0.1$. The hardening rates between the relaxed and the non-relaxed models are within 10\% of each other and is not affected by the initial eccentricity of the orbit. This indicates that the hardening rate is dependent on morphology of the merger product rather than the density of the central cusp. For the mildly eccentric orbits ({\tt\string ecc\_1}), we observe that the hardening rates are lower than that in the circular case whereas in the highly eccentric orbit models ({\tt\string ecc\_2}) demonstrate a larger hardening rate. This is due to the differences in the structure of the final merger product.

 The presence of a relaxed cusp affects the evolution of eccentricity. The relaxed cusp leads to a lower eccentricity at binary formation. For {\tt\string r\_q\_0.1\_ecc\_1} the eccentricity at binary formation is almost 0.0 even though the initial orbit was eccentric. Even for highly eccentric initial orbits like in the {\tt\string r\_q\_0.1\_ecc\_2} scenario, we find that the eccentricity at formation is about 0.3. The situation for the non-relaxed models is quite different as {\tt\string nr\_q\_0.1\_ecc\_1} and {\tt\string nr\_q\_0.1\_ecc\_2} models demonstrate eccentricities of 0.19 and 0.55 at binary formation respectively. A This is because, in the presence of the relaxed cusp, the dynamical friction is able to circularize the binary leading to lower eccentricity as also explained in the previous section. A higher initial eccentricity results in a higher eccentricity at binary formation in our models. \cite{Gualandris2022MNRAS.511.4753G} also report similar findings where the eccentricity at binary formation for the models with steeper cusps is systematically lower than that in the models with shallow cusps. 
 
 In the hard binary phase, the non-relaxed models demonstrate a slight growth in eccentricity whereas in the relaxed models, the eccentricity remains roughly constant. This was also observed in \cite{2012ApJ...744...74G} where the authors found that the eccentricity evolution was roughly constant with time in the model with initial eccentricity. {\tt\string nr\_q\_0.1\_ecc\_2} is able to reach a high eccentricity of 0.8 by 10 Myr whereas its relaxed counterpart only reaches 0.3. The eccentricity evolution in this stage is consistent with the findings of \cite{Sesana2010ApJ...719..851S} where the author report that shallower cusps with $q \approx 0.1$ demonstrate growth in eccentricity. The evolution of eccentricity in this phase has important consequences on determination of MBH merger timescales. Binaries that are able to reach high eccentricities can merge order of magnitudes faster than those on circular orbits.  We plan on systematically studying the evolution of eccentricity as a function of the binary mass ratio in  relaxed and non-relaxed models in a future study.

\subsection{GW Emission from SMBH Binaries}

\begin{figure*}
\includegraphics[width=1.0\textwidth]{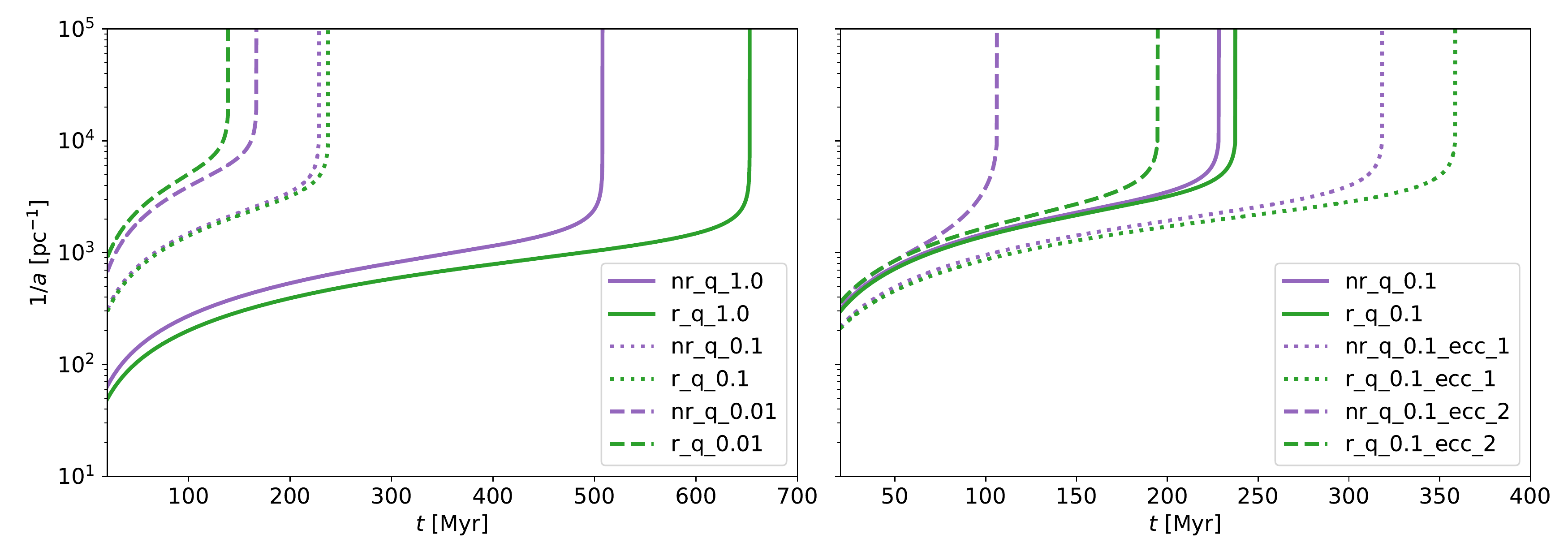}
\caption{The evolution of the inverse-semi major axis as a function of time in the GW dominated phase for different mass-ratios. Left: Evolution and coalescence in models with circular orbits. Right: same but with eccentric models. The evolution is carried by taking the results from the simulations (solid lines) and evolving them semi-analytically (dashed lines) using the Peters \citep{1964PhRv..136.1224P} equation. In the circular models (left), we find that the merger timescales for the non-relaxed models are smaller than their relaxed counterparts for $q=1.0,0.1$ and opposite for $q=0.01$. In the eccentric models (right), we find that the binaries in the non-relaxed cusps always merge faster. All models merge within a Hubble time making NSCs a promising source of GWs.}
\label{fig:gw_emission_all}
\end{figure*}

We follow the evolution of the MBH binaries into the GW coalescence phase semi-analytically after the simulations have been stoppedat $t=10$ Myr. To do so, we assume that the evolution of the inverse semi major axis is constant, i.e. that the hardening due to stellar scattering takes place in the full loss cone limit. This has been the strategy adopted in previous studies \citep[e.g.,][]{2012ApJ...744...74G, 2020MNRAS.493.3676O}. Under this assumption, 

\begin{equation}
    \frac{d}{dt} \left( \frac{1}{a} \right)_{*} = s
    \rightarrow \left( \frac{da}{dt} \right)_{*} = -a^{2}(t)s
\end{equation}
where $s$ is a constant and can be figured out from the inverse semi-major axis data by fitting a straight line through it and measuring its slope. The overall evolution of $a$ can be written as follows
\begin{equation}
    \frac{da}{dt} = \frac{da}{dt} \bigg|_{GW} + \frac{da}{dt} \bigg|_{*}
\end{equation}

\cite{1964PhRv..136.1224P} provides the rate of change of the orbital elements due to the emission of GW. The rate of change of semi-major axis and eccentricity are given as a set of coupled differential equations
\begin{gather}
    \frac{da}{dt} = -\frac{64}{5}\beta \frac{F(e)}{a^{3}} \\
    \frac{de}{dt}  = -\frac{304}{15}\beta \frac{e G(e)}{a^{4}}
\end{gather}
where
\begin{gather}
    \beta = \frac{G^3}{c^5}\left(M_{1}M_{2}\left(M_{1} + M_{2}\right)\right) \\
    G(e) = (1-e^{2})^{-5/2}\left(1+\frac{121}{304}e^2\right) \\
    F(e) = (1-e^2)^{-7/2}\left(1+\frac{73}{24}e^2+\frac{37}{96}e^4\right) .
\end{gather}

We numerically solve the coupled differential equations for both $a$ and $e$ using the orbital elements obtained at end of the $N$-body integration. We do not take into account the growth of eccentricity due to stellar scattering while solving the differential equations. 

From Figure \ref{fig:gw_emission_all}, we find that in all cases with $q \geq 0.1$, the MBH binary in the non-relaxed models merge faster. We find that the {\tt\string nr\_q\_1.0} model undergoes coalescence almost 23\% faster compared to the {\tt\string r\_q\_1.0} model. This is due to the fact that the binary separation itself is lower in the non-relaxed case along with the fact that the scattering rate is also larger because of the reason explained in the previous section. In the $q=0.01$ case, the binary in the relaxed model is at a smaller separation and hardens faster resulting in almost 15\% faster coalescence compared to its non-relaxed counterpart. Since all of our models are roughly circular by the end of the integration time, we expect the timescales presented in \ref{fig:gw_emission_all} to be the upper-limit on the GW merger timescales. 

Interestingly, the coalescence timescale for {\tt\string r\_q\_1.0} and {\tt\string nr\_q\_1.0} simulations is comparable to the relaxation timescale of those systems. It is reasonable to expect the erosion of the non-spherical nature of the system over such timescales in addition to the occurrence of mass-segregation which can affect the hardening rate at later stages for equal mass ratio binaries. However, this is beyond the scope of this study but presents a novel area that merits further investigation in future studies.

We know that the coalescence timescale of the binary depends sensitively on the eccentricity. As such, we expect the binary in the non-relaxed eccentric models to merge faster. Figure \ref{fig:gw_emission_all} shows that this is indeed the case. The MBH binary in {\tt\string nr\_q\_0.1\_ecc\_1} merges by 320 Myr whereas its relaxed counterpart takes around 360 Myr. The effect is stark when considering binaries on highly eccentric orbits. {\tt\string nr\_q\_0.1\_ecc\_2} merges within 95 Myr but {\tt\string r\_q\_0.1\_ecc\_2} takes $2 \times$ longer. Binaries that form in very eccentric orbits thus more efficiently merge in galaxies where the central densities are lower.  Our results are in line with those found by \cite{Gualandris2022MNRAS.511.4753G} where the authors found that models with shallower slopes are more efficient at merging binaries. Our results also underscore the importance of including a mass-spectrum in $N$-body simulations as it can affect collisional relaxation in the NSC which in turn can affect the coalescence timescales.

\subsection{Core Scouring}

 As the binary hardens, it displaces particles from the cusp. The effects of the binary can be strong enough to disrupt the cusp entirely and create a flat core. To understand the effects of the hardening of the binary on the particles, we plot the density profile of both the MS and BH particles in the merged system at different points of hardening. Due to computational limitations, we study the effects up until the time the binary hardens to a semi-major axis $a_{\rm h} /5$ for {\tt\string r\_q\_1.0} and {\tt\string r\_q\_0.1} and upto $\sim a_{\rm h}/2$ for {\tt\string r\_q\_0.01}. The latter model is extremely computationally intensive to evolve longer because of the formation of stable multiple systems in the cusp. Improving the integration scheme to handle secular systems more efficiently \citep{Rantala2022arXiv221002472R} can alleviate this issue.   

In the $q=1.0$ scenario, we find that the inner cusp is completely disrupted and a large flat core is produced as the binary hardens from $a_{\rm h}$ to $a_{\rm h} / 5$. This is seen in the density profiles of both the MS and BH particles. The effective density of the core is $\sim 10^{5} M_{\odot}  \rm{pc}^{-3}$ signalling that the effects of the binary were so strong that the NSC itself was partially disrupted. 

The situation is different for lower mass ratio binaries. For $q=0.1$, the cusp of MS particles is partially disrupted as the binary hardens from $a_{\rm h}$ to $a_{\rm h}/3$ but further disruption is not seen with more hardening. The original cusp is not retained. In fact, for the MS particles, the density profile of the merged system has a faint $\gamma_{\rm MS} = 0.5$ inner slope. For the BH particles, we find a faint $\gamma_{\rm BH}=0.7$ slope. For $q=0.01$, the effects are even more minuscule.  However, since the binary could not be evolved to $a_{\rm h}/5$ due to computational limitations, we exclude it from this analysis.   

The partial retention of the cusp has implications on the regrowth of the Bahcall-Wolf cusp post MBH binary coalescence. Whereas the time required to achieve the collisionally relaxed state for $q=1.0$ may exceed the Hubble time because of the presence of a flat core in both the MS and the BH particles \citep{Merritt2010ApJ...718..739M}, the same cannot be said for $q=0.1,0.01$ which will have faster regrowth.  Crucially, we would need to understand how regrowth inter-plays with the galaxy geometry post merger. As such, $N$-body simulations are required to quantify the exact amount of time required for the regrowth of cusps. 

Since Extreme Mass Ratio Inspiral (EMRI) rates are usually extrapolated under the assumption of a Bahcall-Wolf cusp at the center, we would expect EMRIs to arise out of galaxies that have undergone mergers with lower mass ratios. However, multiple mergers even with lower mass ratios can lead to the formation of a core and the exact number of mergers leading to the formation of a core as a function of $q$ requires further studies. The time dependent rate of EMRIs post merger would be interesting to understand as well and we plan on exploring this in future studies.

How does the disruption of the cusp affect the velocity distribution? To analyze that, we plot the velocity anisotropy parameter for the {\tt\string r\_q\_1.0} model in Figure \ref{fig:anisotropy} during different stages of hardening. The velocity anisotropy parameter is defined as
\begin{equation}
    \beta = 1 - \frac{\sigma_{\rm t}^2}{2 \sigma_{\rm r}^2}
\end{equation}
where $\sigma_{\rm r}$ is the radial velocity dispersion and $\sigma_{\rm t}$ is the tangential velocity dispersion.
We find that as the binary hardens, the velocity profile, which was initially isotropic, becomes tangentially biased . This is caused because the MBH binary preferentially ejects particles on radial orbits. The anisotropy parameter can act as an observational evidence for the presence of an MBH binary because of this reason. This was also noted in previous studies like \cite{2006ApJ...648..890M}. 
\begin{figure*}

\centering
\begin{subfigure}[b]{\linewidth}
\includegraphics[width=0.97\linewidth]{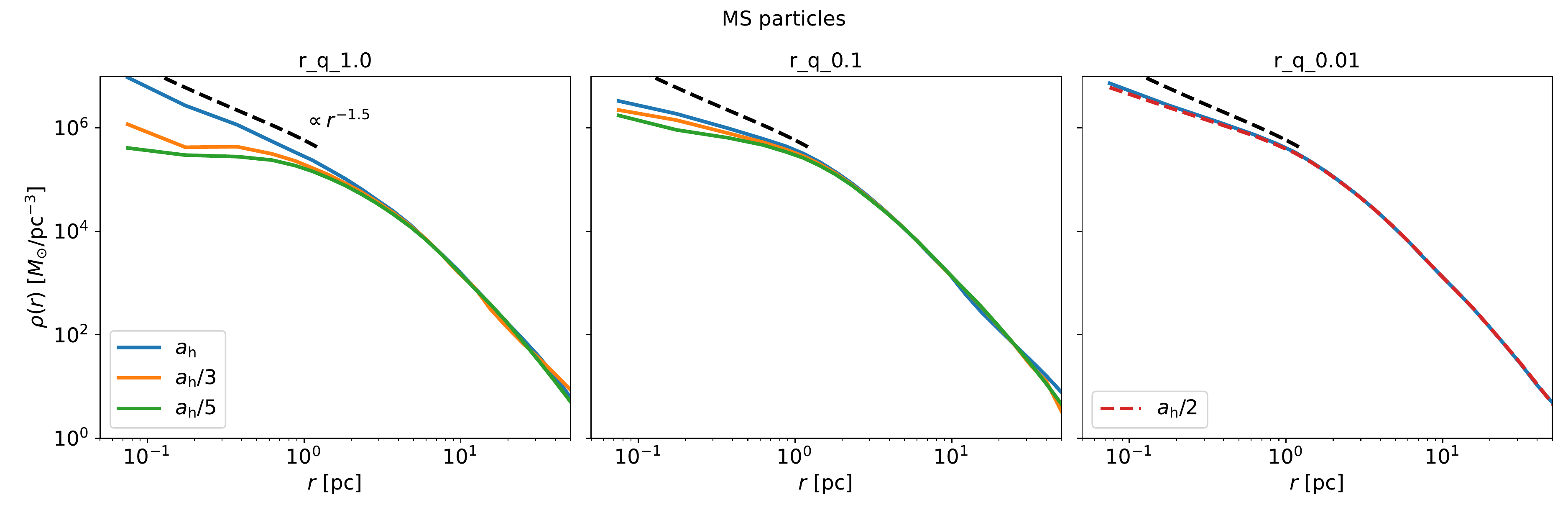}
\end{subfigure}

\begin{subfigure}[b]{\linewidth}
\includegraphics[width=0.97\linewidth]{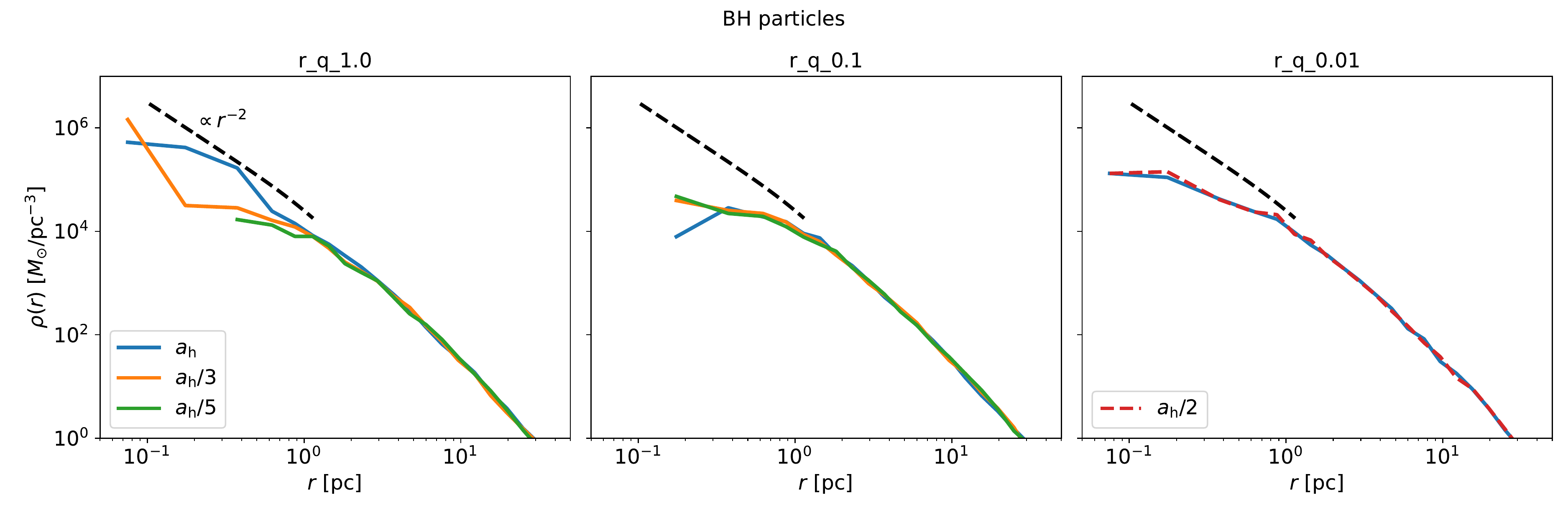}
\end{subfigure}

\caption{The density of particles $\rho$ presented as a function of the distance $r$ from the center of mass of the binary at different points in hardening for {\tt\string r\_} simulations. The initial cusp is also presented for comparison. Top: Density of MS particles. We can see that for $q=1.0$ as the binary hardens, a core is formed. This is not observed for $q=0.1$. Bottom: Density of BH particles. Similar observations are noted in this case.}
\label{fig:core_scouring_all}
\end{figure*}

\begin{figure}
    \centering
    \includegraphics[width=0.5\textwidth]{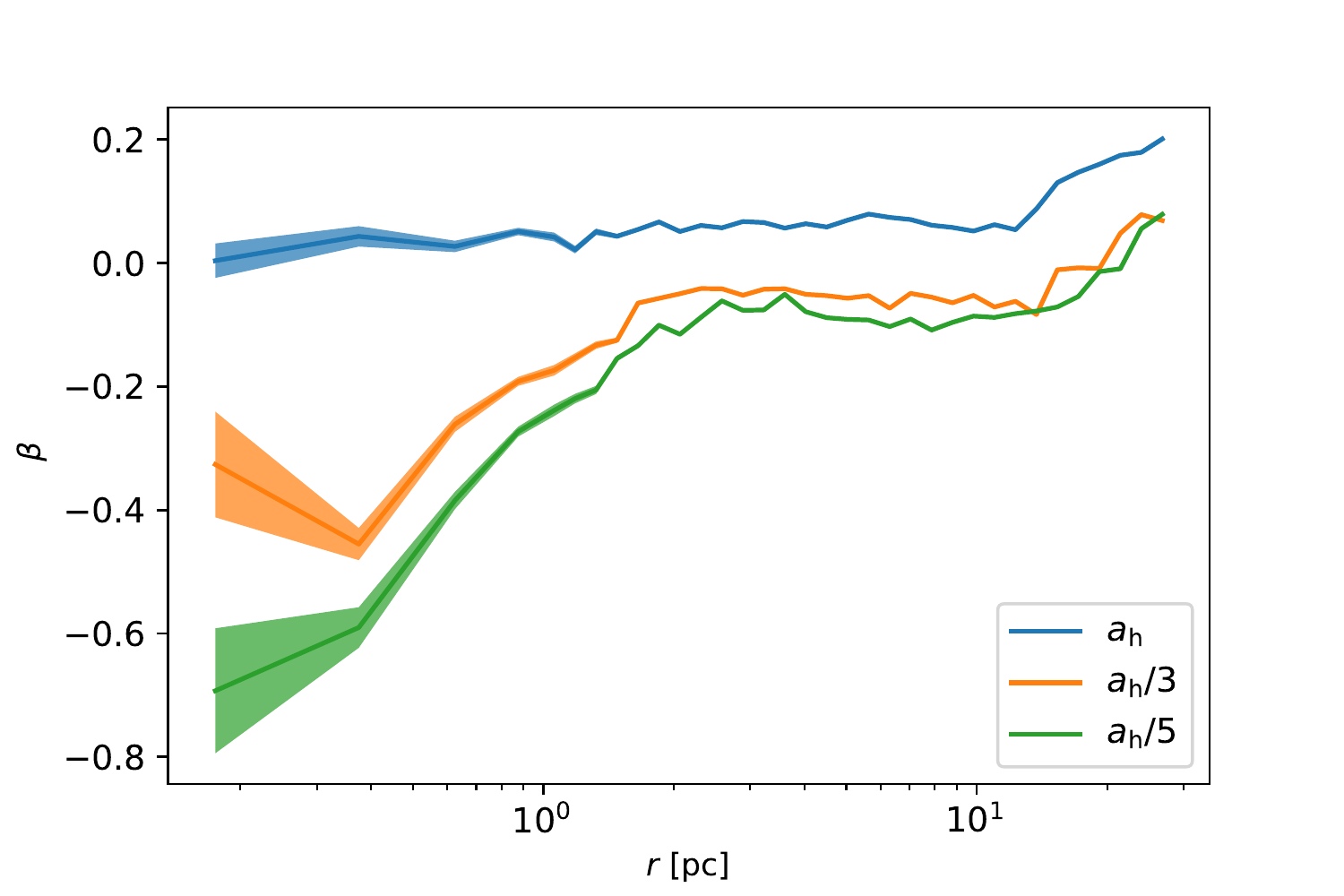}
    \caption{The evolution of the velocity anisotropy for {\tt\string r\_q\_1.0} model. The anisotropy parameter ($\beta$) is plotted as a function of the distance from the binary center($r$). The shaded portions denote the standard error in calculating the anisotropy parameter. The models start initially with an isotropic distribution of velocity. As the binary hardens, it preferentially ejects particles with radial velocities producing a tangentially biased velocity structure near the MBH binary.}
    \label{fig:anisotropy}
\end{figure}

\section{Discussion} \label{sec:discussion}
\subsection{Impact of stochasticity}

As there is inherent stochasticity involved while generating $N$-body samples, one may be curious as to whether the results mentioned in the previous sections are robust and reproducible. To understand the effect of stochasticity on the results, we generate four additional statistically independent merger models for the {\tt\string r\_q\_1.0} and {\tt\string nr\_q\_1.0} simulations. To save computational resources, we only perform the simulations for the $q=1.0$ models up to a termination time of 5 Myr with a resolution of $N \sim 4 \times 10^5$. This should be sufficient to study any discrepancies in the three phases of evolution despite the lower resolution. The impact of resolution for higher mass ratio binaries is minimal and is also further discussed in the next section and appendix \ref{appendix:resolution}.

\begin{figure}
    \centering
    \includegraphics[width=0.5\textwidth]{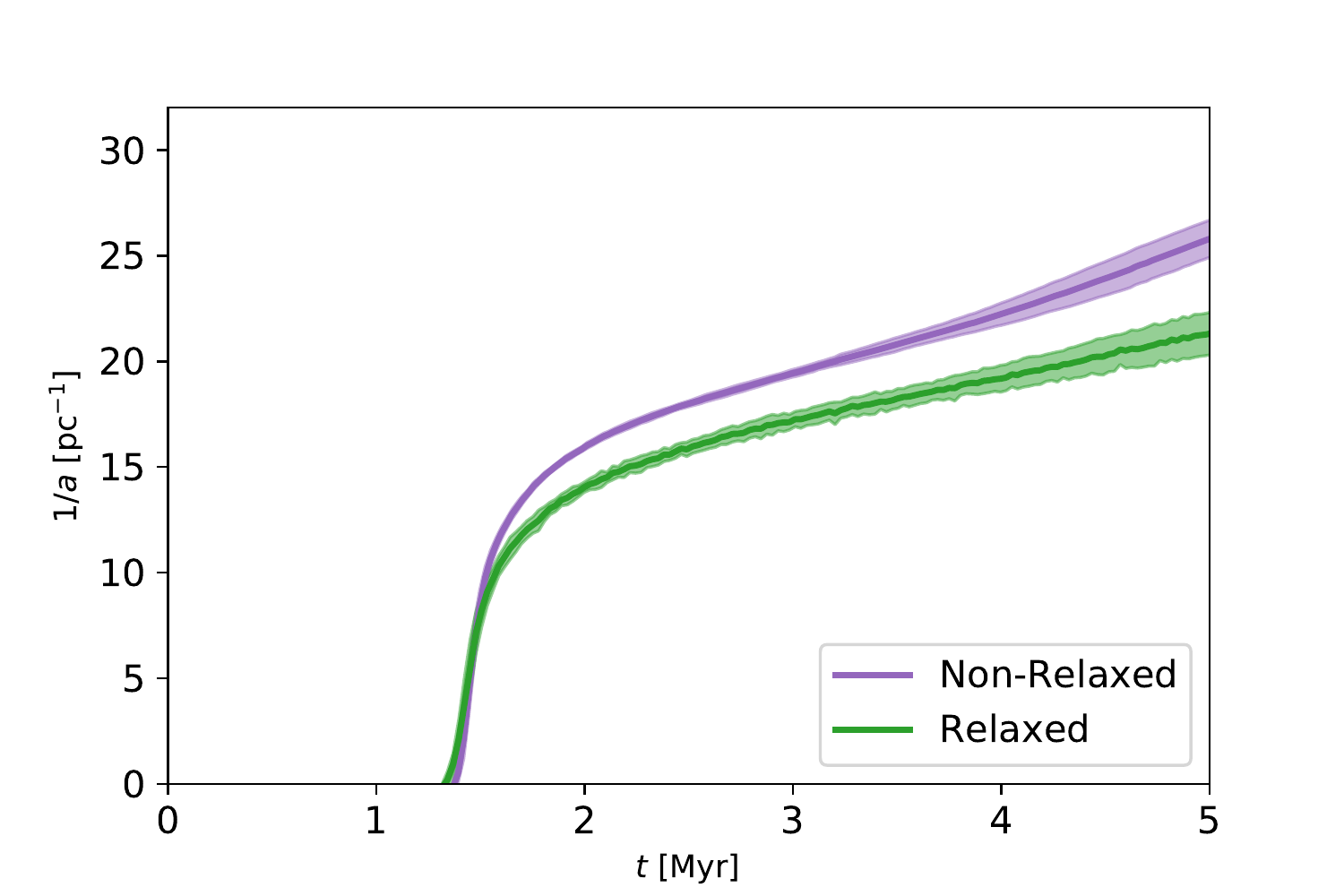}
    \caption{The evolution of the inverse semi-major axis $1/a$ as a function of time $t$ for five independent realizations of {\tt\string nr\_q\_1.0} and {\tt\string r\_q\_1.0} simulations but with a resolution of $N \sim 4 \times 10^5$ for computational constraints. The solid lines represent the mean value of the inverse semi-major axis and the shaded region represents the standard deviation from five simulations. We find that our results are robust. This implies that the discrepancies in the results arise out of physical rather than numerical reasons.}
    \label{fig:robust}
\end{figure}

In Figure \ref{fig:robust}, we plot the mean and standard deviation of the evolution of the inverse semi-major axis from the five simulations. We find that the results from the original simulations are reproducible and the stochastic scatter for the evolution of the separation and semi-major axis is low. This indicates that the differences in the results of the non-relaxed and relaxed models arise out of physical and not numerical reasons. 

We caution the reader that such agreement might not be present for other orbital elements like eccentricity. Previous studies like \cite{Nasim2020MNRAS.497..739N} have highlighted this issue and found that it stems from insufficient numerical resolution. The stochasticity of the stellar encounters with the MBH binary affect eccentricty more leading to larger scatter among random realizations. This can potentially affect GW merger timescales since they are sensitive to the eccentricity of the binary. However, the resolution used in this work is sufficient to study the evolution of the semi-major axis and separation. 

\subsection{Comparison with previous studies} \label{subsec:compare_prev_studies}

 \cite{2012ApJ...744...74G} and \cite{Khan2018A&A...615A..71K} used a non-uniform mass function to model the galaxy mergers. The resolution used in our simulations is lower compared to that in \cite{2012ApJ...744...74G} but is comparable to that used in \cite{Khan2018A&A...615A..71K}.  Although our set up is quite different compared to \cite{2012ApJ...744...74G} we find qualitatively similar results. Comparing Figure 7 from \cite{2012ApJ...744...74G} to Figure \ref{fig:lagrangian_main_seg} in our study, we observe that for the expansion of Lagrangian radius is quite sudden for mass ratios between $0.1-1.0$. Due to the energy injected into the cusp by the MBH binary there is an expansion in the orbits of stars leading to the destruction of the cusp as explained in the previous sections. Our work also suggests consistency with the evolution of the density profile. Although not directly comparable, we find that the  effect of binary leading to core scouring presented for $q=1/3$ in \cite{2012ApJ...744...74G} lies between results from our $q=0.1,1.0$ simulations.

Our results are in contrast with those presented in \cite{Khan2018A&A...615A..71K} where the authors found that the hardening rates in mass segregated cases were significantly higher than those in the non-mass segregated cases, which was not found in our case. The discrepancy could result from the usage of a different initial mass function in \cite{Khan2018A&A...615A..71K} or the usage of a different mass ratio of the binary. The authors admit that the effects of relaxation in their work could be enhanced because of the lower resolution used while the galaxy itself is relaxing prior to the merger. This effect would be negligible in our case since the collisionally relaxed models have been generated from Fokker-Planck simulation rather than $N$-body simulations. It could also occur from different galaxy merger shapes. The interplay between triaxiality and mass segregation has not been properly studied and presents a further avenue of research.

We find similar results compared to \cite{2020MNRAS.493.3676O}. This verifies that the effect of NSCs in accelerating the orbital decline of MBH binaries is robust as indicated by \cite{2020MNRAS.493.3676O}. Our models are characterized by an initial shallow cusp (non-relaxed) or a dense cusp (relaxed), so the results are not directly comparable. The upper limit on the coalescence timescale in our simulations is $\sim 700$ Myr for the $q=1.0$ models whereas that found in \cite{2020MNRAS.493.3676O} is $\sim 5$ Gyr. This could be because the presence of a cusp rather than a core leads to a more efficient hardening. However, \cite{Khan2021MNRAS.508.1174K} found coalescence timescales similar to ours for large mass-ratio mergers and attributed  the discrepancy is  to better resolution of the three-body scattering process with higher mass resolution. Nevertheless, we find that the merger timescale depends on the mass ratio of the binary with $q=0.01$ mass ratio binaries merging in $\sim 150-170$ Myr. Interestingly, this is in contrast to the upper limit to the timescales found in \cite{2020MNRAS.493.3676O}. 

We noted that the timescales for lower mass ratio binaries ($q \leq 10^{-2})$ is quite sensitive to the resolution. Lower resolution simulations with $N \sim 4\times 10^5$ (see appendix \ref{appendix:resolution}) resulted in merger timescales of $\sim 90-100$ Myr for the $q=0.01$ merger simulations. This is almost half the merger timescale found using the higher resolution simulations with $N \sim 1.32 \times 10^6$. As the mass ratio of the binary was increased, the discrepancies between the lower and the higher resolution simulations decreased. This is in contrast to the findings of \cite{Preto2011ApJ...732L..26P} where the authors noted that in MBH binaries formed in galactic mergers, the triaxiality of the non-spherical merger product ensured that the hardening rate is independent of $N$. However, our results seem to be in line with that presented in \cite{Vasilev2015ApJ...810...49V} where the authors found that in triaxial galaxies the hardening rate asymptotically reaches a fixed value as the resolution is increased. According to \cite{Vasilev2014ApJ...785..163V} collisional effects account for a non-trivial portion of the hardening rate and cannot be neglected while considering the hardening rates of MBH binaries. Since NSCs are collisional systems, it highlights the importance of resolving collisional effects properly for a proper determination of LISA timescales for MBH binaries, especially for $ q \approx 0.01$. This would require the usage of even higher particle numbers, even for simulations where the initial orbit is circular. To determine the minimum resolution required, we need to perform more simulations with varying $N$, which is beyond the scope of this study.   We plan on investigating the asymptotic limits of hardening rates as a function of the resolution in future studies. We would like to stress, however, that the binaries merge efficiently well within the Hubble time in line with the conclusions made in \cite{2020MNRAS.493.3676O} about NSCs being potentially important sources for LISA detections.

\section{Conclusions} \label{sec:conclusions}
MBH binaries are touted to be one of the most important sources of GW signals detectable by future generation of GW detectors like LISA. The presence of NSCs surrounding the MBHs can accelerate their evolution to the hard-binary and therefore, the GW emission phase. However, the dynamics of the binary can be quite sensitive to the composition of the NSC. The presence of a mass spectrum can lead to mass segregation since the NSC is a collisional system and can affect the evolution of the binary in non-intuitive ways. 

In this work we have explored the effects of mergers of collisionally relaxed NSC on the dynamics of MBHs embedded in them. Using a suite of $N$-body simulations, we have demonstrated the non-intuitive ways in which a collisionally relaxed nuclei with a two-component initial mass function can affect the overall dynamics depending on the mass ratio of the binary. For simplicity, we considered the mass of the lighter objects to be $1 M_{\odot}$ representing stars, white dwarfs, and neutron stars and the mass of heavier objects to be $10 M_{\odot}$ representing stellar mass black holes.

Through the usage of a Fokker-Planck code, we evolved the NSCs to the collisionally relaxed state under the presence of an MBH at the center. We then set up the mergers with different MBH mass ratios and for comparison, also evolved mergers with non-relaxed NSCs. 

During the three stages of evolution, we found that the dynamics during the pre-binary phase is similar amongst simulations, even with different mass ratios and is consistent with results from \cite{2020MNRAS.493.3676O}. However, due to changes in the density profile between relaxed and non-relaxed systems, differences arise during the combined phase.

The presence of a heavier mass species leads to a decline in the density profile of the lighter species for all radii greater than $\sim 0.1 d_{\rm infl}$ within the sphere of influence of the primary. As a result, in larger mass ratio binaries, the binary is able to settle at a lower separation in the non-relaxed models after the combined phase compared to the relaxed models. However, this trend slowly changes with the mass-ratio of the binary as the scattering efficiency of the binary increases with the decrease in mass-ratio and the binary is able to lose more energy by scattering the particles that are more tightly bound to the MBHs. For $q=0.1$, we find that the evolution in the relaxed and non-relaxed models are similar and for $q=0.01$, we find that the presence of the denser cusp actually accelerated the evolution of the binary to the hard binary stage.

In the hard binary stage, the evolution is similar amongst non-relaxed and relaxed models. The binaries harden at a fixed rate consistent with previous studies on NSC and galaxy mergers. This indicates that even when the cusp is disrupted less, the primary mode of evolution is collisionless where the loss-cone of the binary is populated by stars on centrophillic orbits. This is driven by the shape of the merger product rather than by any relaxation effects. However, we do find that the hardening rate depends on the particle number $N$ in contrast with some previous studies. This underscores the importance of accounting for properly accounting for collisional effects as they plan a non-trivial role in the evolution of MBH binaries in collisional systems like NSCs.

Crucially, we find that the relaxed cusp plays a big role in the determination of the eccentricity at the binary binding and hardening stages. Non-relaxed cusps show higher eccentricities at binding stage and show growth in eccentricity which is absent in relaxed cusps. The eccentricity ``suppression'' is quite large in the relaxed models and binaries on highly eccentric initial orbits can merge almost $2\times$ faster in non-relaxed cusps.

The expected GW coalescence time for all of our models (relaxed and non-relaxed) is significantly less than the Hubble time, making merging NSCs a promising GW source. In addition, we find that the expected coalescence time for the non-relaxed models is lower than their relaxed counterparts for all models with $q > 10^{-2}$. The eccentricity evolution of the MBH binary is strongly affected by the initial density profile of the NSCs with relaxed models exhibiting lower eccentricity. This underscores the importance of properly modeling the initial conditions of the NSC including the usage of a realistic mass-spectrum. Generation of LISA waveforms will require proper modeling of the environment surrounding the binary.

While our initial conditions are idealized in the sense that only two mass species are used, we demonstrate the necessity of modeling NSCs with multiple mass species as collisional relaxation can often evolve the density and mass profiles in non-intuitive ways which can affect the dynamics of the MBH binary. Our simulations open up avenues of further exploration with regards to IMBH-MBH mergers and the impact of triaxiality on mass segregation. We also demonstrate the effectiveness of {\tt\string Taichi} at handling problems of this resolution and scale. {\tt\string Taichi} is among the first $N$-body codes built with a fourth-order symplectic integrator with time symmetric step solver and regularization. Nevertheless, some of the simulations from this work clearly show that directly integrating over many orbits of the hard binaries can be computationally demanding. A proper treatment of stable hierarchical systems \citep[e.g.,][]{Wang2020MNRAS.493.3398W, Rantala2022arXiv221002472R} 
is worth the investment. With further improvements, {\tt\string Taichi} presents an effective method to simulate galaxy mergers, MBH binaries and simulations of NSCs.

\section*{Acknowledgements}

We would like to thank the referee for suggesting valuable
changes to improve the paper. We thank Eugene Vasilev for his help with {\tt\string Agama} and {\tt\string Phaseflow}. We thank Yihan Wang for helping us with implementing regularization in Taichi using {\tt\string SpaceHub}. We thank Manuel Arca Sedda for valuable comments on the work. We acknowledge the usage of the Vera cluster, which is supported by the McWilliams Center for Cosmology and Pittsburgh Supercomputing Center, and the Bridges-2 supercomputer which is maintained by the Pittsburgh Supercomputing Center. We acknowledge computational support from XSEDE grants PHY210042P and PHY200058P. QZ acknowledge the financial support through the McWilliams Post-doctoral Fellowship. GO was supported by the Fundamental Research Fund for Chinese Central Universities (Grant No. NZ2020021) and the Waterloo Centre for Astrophysics Fellowship. CR was supported by NSF Grant AST-2009916 and a New Investigator Research Grant from the Charles E.~Kaufman Foundation. HT acknowledges support from NASA grants 80NSSC22K0722, 80NSSC22K0821, and NSF grant PHY2020295.

\section*{Data Availability}

The data from the $N$-body simulations, and the version of
{\tt\string Taichi} used for this work, are available upon reasonable requests. Please contact the authors if you would be interested in using a development version.



\bibliographystyle{mnras}

\begin{thebibliography}{}
\makeatletter
\relax
\def\mn@urlcharsother{\let\do\@makeother \do\$\do\&\do\#\do\^\do\_\do\%\do\~}
\def\mn@doi{\begingroup\mn@urlcharsother \@ifnextchar [ {\mn@doi@}
  {\mn@doi@[]}}
\def\mn@doi@[#1]#2{\def\@tempa{#1}\ifx\@tempa\@empty \href
  {http://dx.doi.org/#2} {doi:#2}\else \href {http://dx.doi.org/#2} {#1}\fi
  \endgroup}
\def\mn@eprint#1#2{\mn@eprint@#1:#2::\@nil}
\def\mn@eprint@arXiv#1{\href {http://arxiv.org/abs/#1} {{\tt arXiv:#1}}}
\def\mn@eprint@dblp#1{\href {http://dblp.uni-trier.de/rec/bibtex/#1.xml}
  {dblp:#1}}
\def\mn@eprint@#1:#2:#3:#4\@nil{\def\@tempa {#1}\def\@tempb {#2}\def\@tempc
  {#3}\ifx \@tempc \@empty \let \@tempc \@tempb \let \@tempb \@tempa \fi \ifx
  \@tempb \@empty \def\@tempb {arXiv}\fi \@ifundefined
  {mn@eprint@\@tempb}{\@tempb:\@tempc}{\expandafter \expandafter \csname
  mn@eprint@\@tempb\endcsname \expandafter{\@tempc}}}

\bibitem[\protect\citeauthoryear{{Alexander} \& {Hopman}}{{Alexander} \&
  {Hopman}}{2009}]{2009ApJ...697.1861A}
{Alexander} T.,  {Hopman} C.,  2009, \mn@doi [\apj]
  {10.1088/0004-637X/697/2/1861}, \href
  {https://ui.adsabs.harvard.edu/abs/2009ApJ...697.1861A} {697, 1861}

\bibitem[\protect\citeauthoryear{{Amaro-Seoane} et~al.,}{{Amaro-Seoane}
  et~al.}{2017}]{2017arXiv170200786A}
{Amaro-Seoane} P.,  et~al., 2017, arXiv e-prints, \href
  {https://ui.adsabs.harvard.edu/abs/2017arXiv170200786A} {p. arXiv:1702.00786}

\bibitem[\protect\citeauthoryear{{Antonini}}{{Antonini}}{2014}]{Antonini2014ApJ...794..106A}
{Antonini} F.,  2014, \mn@doi [\apj] {10.1088/0004-637X/794/2/106}, \href
  {https://ui.adsabs.harvard.edu/abs/2014ApJ...794..106A} {794, 106}

\bibitem[\protect\citeauthoryear{{Arca-Sedda} \& {Gualandris}}{{Arca-Sedda} \&
  {Gualandris}}{2018}]{ArcaSedda2018MNRAS.477.4423A}
{Arca-Sedda} M.,  {Gualandris} A.,  2018, \mn@doi [\mnras]
  {10.1093/mnras/sty922}, \href
  {https://ui.adsabs.harvard.edu/abs/2018MNRAS.477.4423A} {477, 4423}

\bibitem[\protect\citeauthoryear{{Baes}, {Dejonghe}  \& {Buyle}}{{Baes}
  et~al.}{2005}]{Baes2005A&A...432..411B}
{Baes} M.,  {Dejonghe} H.,   {Buyle} P.,  2005, \mn@doi [\aap]
  {10.1051/0004-6361:20041907}, \href
  {https://ui.adsabs.harvard.edu/abs/2005A&A...432..411B} {432, 411}

\bibitem[\protect\citeauthoryear{{Bahcall} \& {Wolf}}{{Bahcall} \&
  {Wolf}}{1976}]{1976ApJ...209..214B}
{Bahcall} J.~N.,  {Wolf} R.~A.,  1976, \mn@doi [\apj] {10.1086/154711}, \href
  {https://ui.adsabs.harvard.edu/abs/1976ApJ...209..214B} {209, 214}

\bibitem[\protect\citeauthoryear{{Bahcall} \& {Wolf}}{{Bahcall} \&
  {Wolf}}{1977}]{1977ApJ...216..883B}
{Bahcall} J.~N.,  {Wolf} R.~A.,  1977, \mn@doi [\apj] {10.1086/155534}, \href
  {https://ui.adsabs.harvard.edu/abs/1977ApJ...216..883B} {216, 883}

\bibitem[\protect\citeauthoryear{{Bartko} et~al.,}{{Bartko}
  et~al.}{2010}]{2010ApJ...708..834B}
{Bartko} H.,  et~al., 2010, \mn@doi [\apj] {10.1088/0004-637X/708/1/834}, \href
  {https://ui.adsabs.harvard.edu/abs/2010ApJ...708..834B} {708, 834}

\bibitem[\protect\citeauthoryear{{Begelman}, {Blandford}  \& {Rees}}{{Begelman}
  et~al.}{1980}]{Begelman1980Natur.287..307B}
{Begelman} M.~C.,  {Blandford} R.~D.,   {Rees} M.~J.,  1980, \mn@doi [\nat]
  {10.1038/287307a0}, \href
  {https://ui.adsabs.harvard.edu/abs/1980Natur.287..307B} {287, 307}

\bibitem[\protect\citeauthoryear{{Berczik}, {Merritt}, {Spurzem}  \&
  {Bischof}}{{Berczik} et~al.}{2006}]{Berczik2006ApJ...642L..21B}
{Berczik} P.,  {Merritt} D.,  {Spurzem} R.,   {Bischof} H.-P.,  2006, \mn@doi
  [\apjl] {10.1086/504426}, \href
  {https://ui.adsabs.harvard.edu/abs/2006ApJ...642L..21B} {642, L21}

\bibitem[\protect\citeauthoryear{{Bogdanovi{\'c}}}{{Bogdanovi{\'c}}}{2015}]{2015ASSP...40..103B}
{Bogdanovi{\'c}} T.,  2015, in Gravitational Wave Astrophysics. p.~103
  (\mn@eprint {arXiv} {1406.5193}), \mn@doi{10.1007/978-3-319-10488-1\_9}

\bibitem[\protect\citeauthoryear{{Chabrier}}{{Chabrier}}{2005}]{Chabrier2005ASSL..327...41C}
{Chabrier} G.,  2005, in {Corbelli} E.,  {Palla} F.,   {Zinnecker} H.,  eds,
  Astrophysics and Space Science Library Vol. 327, The Initial Mass Function 50
  Years Later. p.~41 (\mn@eprint {arXiv} {astro-ph/0409465}),
  \mn@doi{10.1007/978-1-4020-3407-7\_5}

\bibitem[\protect\citeauthoryear{Cheng, Greengard  \& Rokhlin}{Cheng
  et~al.}{1999}]{cheng1999fast}
Cheng H.,  Greengard L.,   Rokhlin V.,  1999, Journal of computational physics,
  155, 468

\bibitem[\protect\citeauthoryear{{Chin}}{{Chin}}{1997}]{1997PhLA..226..344C}
{Chin} S.~A.,  1997, \mn@doi [Physics Letters A]
  {10.1016/S0375-9601(97)00003-0}, \href
  {https://ui.adsabs.harvard.edu/abs/1997PhLA..226..344C} {226, 344}

\bibitem[\protect\citeauthoryear{{Chin}}{{Chin}}{2007}]{Chin2007PhRvE..75c6701C}
{Chin} S.~A.,  2007, \mn@doi [\pre] {10.1103/PhysRevE.75.036701}, \href
  {https://ui.adsabs.harvard.edu/abs/2007PhRvE..75c6701C} {75, 036701}

\bibitem[\protect\citeauthoryear{{Chin} \& {Chen}}{{Chin} \&
  {Chen}}{2005}]{Chen2005CeMDA..91..301C}
{Chin} S.~A.,  {Chen} C.~R.,  2005, \mn@doi [Celestial Mechanics and Dynamical
  Astronomy] {10.1007/s10569-004-4622-z}, \href
  {https://ui.adsabs.harvard.edu/abs/2005CeMDA..91..301C} {91, 301}

\bibitem[\protect\citeauthoryear{{Coulaud}, {Fortin}  \& {Roman}}{{Coulaud}
  et~al.}{2008}]{Coulaud2008JCoPh.227.1836C}
{Coulaud} O.,  {Fortin} P.,   {Roman} J.,  2008, \mn@doi [Journal of
  Computational Physics] {10.1016/j.jcp.2007.09.027}, \href
  {https://ui.adsabs.harvard.edu/abs/2008JCoPh.227.1836C} {227, 1836}

\bibitem[\protect\citeauthoryear{{Dehnen}}{{Dehnen}}{1993}]{Dehnen1993MNRAS.265..250D}
{Dehnen} W.,  1993, \mn@doi [\mnras] {10.1093/mnras/265.1.250}, \href
  {https://ui.adsabs.harvard.edu/abs/1993MNRAS.265..250D} {265, 250}

\bibitem[\protect\citeauthoryear{{Dehnen}}{{Dehnen}}{2005}]{2005MNRAS.360..892D}
{Dehnen} W.,  2005, \mn@doi [\mnras] {10.1111/j.1365-2966.2005.09099.x}, \href
  {https://ui.adsabs.harvard.edu/abs/2005MNRAS.360..892D} {360, 892}

\bibitem[\protect\citeauthoryear{{Farr} \& {Bertschinger}}{{Farr} \&
  {Bertschinger}}{2007}]{Farr2007ApJ...663.1420F}
{Farr} W.~M.,  {Bertschinger} E.,  2007, \mn@doi [\apj] {10.1086/518641}, \href
  {https://ui.adsabs.harvard.edu/abs/2007ApJ...663.1420F} {663, 1420}

\bibitem[\protect\citeauthoryear{{Ghez} et~al.,}{{Ghez}
  et~al.}{2008}]{2008ApJ...689.1044G}
{Ghez} A.~M.,  et~al., 2008, \mn@doi [\apj] {10.1086/592738}, \href
  {https://ui.adsabs.harvard.edu/abs/2008ApJ...689.1044G} {689, 1044}

\bibitem[\protect\citeauthoryear{Greengard \& Rokhlin}{Greengard \&
  Rokhlin}{1987}]{greengard1987fast}
Greengard L.,  Rokhlin V.,  1987, Journal of computational physics, 73, 325

\bibitem[\protect\citeauthoryear{{Gualandris} \& {Merritt}}{{Gualandris} \&
  {Merritt}}{2012}]{2012ApJ...744...74G}
{Gualandris} A.,  {Merritt} D.,  2012, \mn@doi [\apj]
  {10.1088/0004-637X/744/1/74}, \href
  {https://ui.adsabs.harvard.edu/abs/2012ApJ...744...74G} {744, 74}

\bibitem[\protect\citeauthoryear{{Gualandris}, {Khan}, {Bortolas}, {Bonetti},
  {Sesana}, {Berczik}  \& {Holley-Bockelmann}}{{Gualandris}
  et~al.}{2022}]{Gualandris2022MNRAS.511.4753G}
{Gualandris} A.,  {Khan} F.~M.,  {Bortolas} E.,  {Bonetti} M.,  {Sesana} A.,
  {Berczik} P.,   {Holley-Bockelmann} K.,  2022, \mn@doi [\mnras]
  {10.1093/mnras/stac241}, \href
  {https://ui.adsabs.harvard.edu/abs/2022MNRAS.511.4753G} {511, 4753}

\bibitem[\protect\citeauthoryear{{Hopman} \& {Alexander}}{{Hopman} \&
  {Alexander}}{2006}]{Hopman2006ApJ...645L.133H}
{Hopman} C.,  {Alexander} T.,  2006, \mn@doi [\apjl] {10.1086/506273}, \href
  {https://ui.adsabs.harvard.edu/abs/2006ApJ...645L.133H} {645, L133}

\bibitem[\protect\citeauthoryear{{Huang}}{{Huang}}{1963}]{Huang1963ApJ...138..471H}
{Huang} S.-S.,  1963, \mn@doi [\apj] {10.1086/147659}, \href
  {https://ui.adsabs.harvard.edu/abs/1963ApJ...138..471H} {138, 471}

\bibitem[\protect\citeauthoryear{{Khan} \& {Holley-Bockelmann}}{{Khan} \&
  {Holley-Bockelmann}}{2021}]{Khan2021MNRAS.508.1174K}
{Khan} F.~M.,  {Holley-Bockelmann} K.,  2021, \mn@doi [\mnras]
  {10.1093/mnras/stab2646}, \href
  {https://ui.adsabs.harvard.edu/abs/2021MNRAS.508.1174K} {508, 1174}

\bibitem[\protect\citeauthoryear{{Khan}, {Holley-Bockelmann}, {Berczik}  \&
  {Just}}{{Khan} et~al.}{2013}]{Khan2013ApJ...773..100K}
{Khan} F.~M.,  {Holley-Bockelmann} K.,  {Berczik} P.,   {Just} A.,  2013,
  \mn@doi [\apj] {10.1088/0004-637X/773/2/100}, \href
  {https://ui.adsabs.harvard.edu/abs/2013ApJ...773..100K} {773, 100}

\bibitem[\protect\citeauthoryear{{Khan}, {Holley-Bockelmann}  \&
  {Berczik}}{{Khan} et~al.}{2015}]{Khan2015ApJ...798..103K}
{Khan} F.~M.,  {Holley-Bockelmann} K.,   {Berczik} P.,  2015, \mn@doi [\apj]
  {10.1088/0004-637X/798/2/103}, \href
  {https://ui.adsabs.harvard.edu/abs/2015ApJ...798..103K} {798, 103}

\bibitem[\protect\citeauthoryear{{Khan}, {Berczik}  \& {Just}}{{Khan}
  et~al.}{2018}]{Khan2018A&A...615A..71K}
{Khan} F.~M.,  {Berczik} P.,   {Just} A.,  2018, \mn@doi [\aap]
  {10.1051/0004-6361/201730489}, \href
  {https://ui.adsabs.harvard.edu/abs/2018A&A...615A..71K} {615, A71}

\bibitem[\protect\citeauthoryear{{Komossa}}{{Komossa}}{2006}]{2006MmSAI..77..733K}
{Komossa} S.,  2006, \memsai, \href
  {https://ui.adsabs.harvard.edu/abs/2006MmSAI..77..733K} {77, 733}

\bibitem[\protect\citeauthoryear{{Komossa}, {Burwitz}, {Hasinger}, {Predehl},
  {Kaastra}  \& {Ikebe}}{{Komossa} et~al.}{2003}]{2003ApJ...582L..15K}
{Komossa} S.,  {Burwitz} V.,  {Hasinger} G.,  {Predehl} P.,  {Kaastra} J.~S.,
  {Ikebe} Y.,  2003, \mn@doi [\apjl] {10.1086/346145}, \href
  {https://ui.adsabs.harvard.edu/abs/2003ApJ...582L..15K} {582, L15}

\bibitem[\protect\citeauthoryear{{Kormendy} \& {Gebhardt}}{{Kormendy} \&
  {Gebhardt}}{2001}]{2001AIPC..586..363K}
{Kormendy} J.,  {Gebhardt} K.,  2001, in {Wheeler} J.~C.,  {Martel} H.,  eds,
  American Institute of Physics Conference Series Vol. 586, 20th Texas
  Symposium on relativistic astrophysics. pp 363--381 (\mn@eprint {arXiv}
  {astro-ph/0105230}), \mn@doi{10.1063/1.1419581}

\bibitem[\protect\citeauthoryear{{Kormendy} \& {Ho}}{{Kormendy} \&
  {Ho}}{2013}]{2013ARA&A..51..511K}
{Kormendy} J.,  {Ho} L.~C.,  2013, \mn@doi [\araa]
  {10.1146/annurev-astro-082708-101811}, \href
  {https://ui.adsabs.harvard.edu/abs/2013ARA&A..51..511K} {51, 511}

\bibitem[\protect\citeauthoryear{{Kroupa}}{{Kroupa}}{2001}]{Kroupa2001MNRAS.322..231K}
{Kroupa} P.,  2001, \mn@doi [\mnras] {10.1046/j.1365-8711.2001.04022.x}, \href
  {https://ui.adsabs.harvard.edu/abs/2001MNRAS.322..231K} {322, 231}

\bibitem[\protect\citeauthoryear{{Luo} et~al.,}{{Luo}
  et~al.}{2016}]{2016CQGra..33c5010L}
{Luo} J.,  et~al., 2016, \mn@doi [Classical and Quantum Gravity]
  {10.1088/0264-9381/33/3/035010}, \href
  {https://ui.adsabs.harvard.edu/abs/2016CQGra..33c5010L} {33, 035010}

\bibitem[\protect\citeauthoryear{{Ma}, {Hopkins}, {Ma},
  {Angl{\'e}s-Alc{\'a}zar}, {Faucher-Gigu{\`e}re}  \& {Kelley}}{{Ma}
  et~al.}{2021}]{Ma2021MNRAS.508.1973M}
{Ma} L.,  {Hopkins} P.~F.,  {Ma} X.,  {Angl{\'e}s-Alc{\'a}zar} D.,
  {Faucher-Gigu{\`e}re} C.-A.,   {Kelley} L.~Z.,  2021, \mn@doi [\mnras]
  {10.1093/mnras/stab2713}, \href
  {https://ui.adsabs.harvard.edu/abs/2021MNRAS.508.1973M} {508, 1973}

\bibitem[\protect\citeauthoryear{{Makino} \& {Aarseth}}{{Makino} \&
  {Aarseth}}{1992}]{Makino1992PASJ...44..141M}
{Makino} J.,  {Aarseth} S.~J.,  1992, \pasj, \href
  {https://ui.adsabs.harvard.edu/abs/1992PASJ...44..141M} {44, 141}

\bibitem[\protect\citeauthoryear{{Makino}, {Hut}, {Kaplan}  \&
  {Sayg{\i}n}}{{Makino} et~al.}{2006}]{Makino2006NewA...12..124M}
{Makino} J.,  {Hut} P.,  {Kaplan} M.,   {Sayg{\i}n} H.,  2006, \mn@doi [\na]
  {10.1016/j.newast.2006.06.003}, \href
  {https://ui.adsabs.harvard.edu/abs/2006NewA...12..124M} {12, 124}

\bibitem[\protect\citeauthoryear{{Maness} et~al.,}{{Maness}
  et~al.}{2007}]{Maness2007ApJ...669.1024M}
{Maness} H.,  et~al., 2007, \mn@doi [\apj] {10.1086/521669}, \href
  {https://ui.adsabs.harvard.edu/abs/2007ApJ...669.1024M} {669, 1024}

\bibitem[\protect\citeauthoryear{{Merritt}}{{Merritt}}{2010}]{Merritt2010ApJ...718..739M}
{Merritt} D.,  2010, \mn@doi [\apj] {10.1088/0004-637X/718/2/739}, \href
  {https://ui.adsabs.harvard.edu/abs/2010ApJ...718..739M} {718, 739}

\bibitem[\protect\citeauthoryear{{Merritt}}{{Merritt}}{2013}]{2013degn.book.....M}
{Merritt} D.,  2013, {Dynamics and Evolution of Galactic Nuclei}

\bibitem[\protect\citeauthoryear{{Merritt} \& {Szell}}{{Merritt} \&
  {Szell}}{2006}]{2006ApJ...648..890M}
{Merritt} D.,  {Szell} A.,  2006, \mn@doi [\apj] {10.1086/506010}, \href
  {https://ui.adsabs.harvard.edu/abs/2006ApJ...648..890M} {648, 890}

\bibitem[\protect\citeauthoryear{{Mikkola} \& {Tanikawa}}{{Mikkola} \&
  {Tanikawa}}{1999}]{1999MNRAS.310..745M}
{Mikkola} S.,  {Tanikawa} K.,  1999, \mn@doi [\mnras]
  {10.1046/j.1365-8711.1999.02982.x}, \href
  {https://ui.adsabs.harvard.edu/abs/1999MNRAS.310..745M} {310, 745}

\bibitem[\protect\citeauthoryear{{Milosavljevi{\'c}} \&
  {Merritt}}{{Milosavljevi{\'c}} \& {Merritt}}{2003}]{Milo2003AIPC..686..201M}
{Milosavljevi{\'c}} M.,  {Merritt} D.,  2003, in {Centrella} J.~M.,  ed.,
  American Institute of Physics Conference Series Vol. 686, The Astrophysics of
  Gravitational Wave Sources. pp 201--210 (\mn@eprint {arXiv}
  {astro-ph/0212270}), \mn@doi{10.1063/1.1629432}

\bibitem[\protect\citeauthoryear{{Mukherjee}, {Zhu}, {Trac}  \&
  {Rodriguez}}{{Mukherjee} et~al.}{2021}]{2021ApJ...916....9M}
{Mukherjee} D.,  {Zhu} Q.,  {Trac} H.,   {Rodriguez} C.~L.,  2021, \mn@doi
  [\apj] {10.3847/1538-4357/ac03b2}, \href
  {https://ui.adsabs.harvard.edu/abs/2021ApJ...916....9M} {916, 9}

\bibitem[\protect\citeauthoryear{{Nasim}, {Gualandris}, {Read}, {Dehnen},
  {Delorme}  \& {Antonini}}{{Nasim} et~al.}{2020}]{Nasim2020MNRAS.497..739N}
{Nasim} I.,  {Gualandris} A.,  {Read} J.,  {Dehnen} W.,  {Delorme} M.,
  {Antonini} F.,  2020, \mn@doi [\mnras] {10.1093/mnras/staa1896}, \href
  {https://ui.adsabs.harvard.edu/abs/2020MNRAS.497..739N} {497, 739}

\bibitem[\protect\citeauthoryear{{Neumayer}, {Seth}  \& {B{\"o}ker}}{{Neumayer}
  et~al.}{2020}]{2020A&ARv..28....4N}
{Neumayer} N.,  {Seth} A.,   {B{\"o}ker} T.,  2020, \mn@doi [\aapr]
  {10.1007/s00159-020-00125-0}, \href
  {https://ui.adsabs.harvard.edu/abs/2020A&ARv..28....4N} {28, 4}

\bibitem[\protect\citeauthoryear{{Ogiya}, {Hahn}, {Mingarelli}  \&
  {Volonteri}}{{Ogiya} et~al.}{2020}]{2020MNRAS.493.3676O}
{Ogiya} G.,  {Hahn} O.,  {Mingarelli} C. M.~F.,   {Volonteri} M.,  2020,
  \mn@doi [\mnras] {10.1093/mnras/staa444}, \href
  {https://ui.adsabs.harvard.edu/abs/2020MNRAS.493.3676O} {493, 3676}

\bibitem[\protect\citeauthoryear{{Omelyan}}{{Omelyan}}{2006}]{Omelyan2006PhRvE..74c6703O}
{Omelyan} I.~P.,  2006, \mn@doi [\pre] {10.1103/PhysRevE.74.036703}, \href
  {https://ui.adsabs.harvard.edu/abs/2006PhRvE..74c6703O} {74, 036703}

\bibitem[\protect\citeauthoryear{{Pelupessy}, {J{\"a}nes}  \& {Portegies
  Zwart}}{{Pelupessy} et~al.}{2012}]{Pelupessy2012NewA...17..711P}
{Pelupessy} F.~I.,  {J{\"a}nes} J.,   {Portegies Zwart} S.,  2012, \mn@doi
  [\na] {10.1016/j.newast.2012.05.009}, \href
  {https://ui.adsabs.harvard.edu/abs/2012NewA...17..711P} {17, 711}

\bibitem[\protect\citeauthoryear{{Peters}}{{Peters}}{1964}]{1964PhRv..136.1224P}
{Peters} P.~C.,  1964, \mn@doi [Physical Review] {10.1103/PhysRev.136.B1224},
  \href {https://ui.adsabs.harvard.edu/abs/1964PhRv..136.1224P} {136, 1224}

\bibitem[\protect\citeauthoryear{{Preto} \& {Amaro-Seoane}}{{Preto} \&
  {Amaro-Seoane}}{2010}]{Preto2010ApJ...708L..42P}
{Preto} M.,  {Amaro-Seoane} P.,  2010, \mn@doi [\apjl]
  {10.1088/2041-8205/708/1/L42}, \href
  {https://ui.adsabs.harvard.edu/abs/2010ApJ...708L..42P} {708, L42}

\bibitem[\protect\citeauthoryear{{Preto}, {Berentzen}, {Berczik}  \&
  {Spurzem}}{{Preto} et~al.}{2011}]{Preto2011ApJ...732L..26P}
{Preto} M.,  {Berentzen} I.,  {Berczik} P.,   {Spurzem} R.,  2011, \mn@doi
  [\apjl] {10.1088/2041-8205/732/2/L26}, \href
  {https://ui.adsabs.harvard.edu/abs/2011ApJ...732L..26P} {732, L26}

\bibitem[\protect\citeauthoryear{{Rantala}, {Naab}  \& {Springel}}{{Rantala}
  et~al.}{2021}]{2021MNRAS.502.5546R}
{Rantala} A.,  {Naab} T.,   {Springel} V.,  2021, \mn@doi [\mnras]
  {10.1093/mnras/stab057}, \href
  {https://ui.adsabs.harvard.edu/abs/2021MNRAS.502.5546R} {502, 5546}

\bibitem[\protect\citeauthoryear{{Rantala}, {Naab}, {Rizzuto}, {Mannerkoski},
  {Partmann}  \& {Lautensch{\"u}tz}}{{Rantala}
  et~al.}{2022}]{Rantala2022arXiv221002472R}
{Rantala} A.,  {Naab} T.,  {Rizzuto} F.~P.,  {Mannerkoski} M.,  {Partmann} C.,
   {Lautensch{\"u}tz} K.,  2022, arXiv e-prints, \href
  {https://ui.adsabs.harvard.edu/abs/2022arXiv221002472R} {p. arXiv:2210.02472}

\bibitem[\protect\citeauthoryear{{Rodriguez-Gomez} et~al.,}{{Rodriguez-Gomez}
  et~al.}{2016}]{RodriguezGomez2016MNRAS.458.2371R}
{Rodriguez-Gomez} V.,  et~al., 2016, \mn@doi [\mnras] {10.1093/mnras/stw456},
  \href {https://ui.adsabs.harvard.edu/abs/2016MNRAS.458.2371R} {458, 2371}

\bibitem[\protect\citeauthoryear{{S{\'a}nchez-Janssen}
  et~al.,}{{S{\'a}nchez-Janssen} et~al.}{2019}]{2019ApJ...878...18S}
{S{\'a}nchez-Janssen} R.,  et~al., 2019, \mn@doi [\apj]
  {10.3847/1538-4357/aaf4fd}, \href
  {https://ui.adsabs.harvard.edu/abs/2019ApJ...878...18S} {878, 18}

\bibitem[\protect\citeauthoryear{{Sesana}}{{Sesana}}{2010}]{Sesana2010ApJ...719..851S}
{Sesana} A.,  2010, \mn@doi [\apj] {10.1088/0004-637X/719/1/851}, \href
  {https://ui.adsabs.harvard.edu/abs/2010ApJ...719..851S} {719, 851}

\bibitem[\protect\citeauthoryear{{Sesana}, {Haardt}  \& {Madau}}{{Sesana}
  et~al.}{2008}]{Sesana2008ApJ...686..432S}
{Sesana} A.,  {Haardt} F.,   {Madau} P.,  2008, \mn@doi [\apj]
  {10.1086/590651}, \href
  {https://ui.adsabs.harvard.edu/abs/2008ApJ...686..432S} {686, 432}

\bibitem[\protect\citeauthoryear{{Vasiliev}}{{Vasiliev}}{2017}]{Vasilev2017ApJ...848...10V}
{Vasiliev} E.,  2017, \mn@doi [\apj] {10.3847/1538-4357/aa8cc8}, \href
  {https://ui.adsabs.harvard.edu/abs/2017ApJ...848...10V} {848, 10}

\bibitem[\protect\citeauthoryear{{Vasiliev}}{{Vasiliev}}{2019}]{Vasilev2019MNRAS.482.1525V}
{Vasiliev} E.,  2019, \mn@doi [\mnras] {10.1093/mnras/sty2672}, \href
  {https://ui.adsabs.harvard.edu/abs/2019MNRAS.482.1525V} {482, 1525}

\bibitem[\protect\citeauthoryear{{Vasiliev}, {Antonini}  \&
  {Merritt}}{{Vasiliev} et~al.}{2014}]{Vasilev2014ApJ...785..163V}
{Vasiliev} E.,  {Antonini} F.,   {Merritt} D.,  2014, \mn@doi [\apj]
  {10.1088/0004-637X/785/2/163}, \href
  {https://ui.adsabs.harvard.edu/abs/2014ApJ...785..163V} {785, 163}

\bibitem[\protect\citeauthoryear{{Vasiliev}, {Antonini}  \&
  {Merritt}}{{Vasiliev} et~al.}{2015}]{Vasilev2015ApJ...810...49V}
{Vasiliev} E.,  {Antonini} F.,   {Merritt} D.,  2015, \mn@doi [\apj]
  {10.1088/0004-637X/810/1/49}, \href
  {https://ui.adsabs.harvard.edu/abs/2015ApJ...810...49V} {810, 49}

\bibitem[\protect\citeauthoryear{{Wang}, {Spurzem}, {Aarseth}, {Nitadori},
  {Berczik}, {Kouwenhoven}  \& {Naab}}{{Wang}
  et~al.}{2015}]{2015MNRAS.450.4070W}
{Wang} L.,  {Spurzem} R.,  {Aarseth} S.,  {Nitadori} K.,  {Berczik} P.,
  {Kouwenhoven} M.~B.~N.,   {Naab} T.,  2015, \mn@doi [\mnras]
  {10.1093/mnras/stv817}, \href
  {https://ui.adsabs.harvard.edu/abs/2015MNRAS.450.4070W} {450, 4070}

\bibitem[\protect\citeauthoryear{{Wang}, {Nitadori}  \& {Makino}}{{Wang}
  et~al.}{2020}]{Wang2020MNRAS.493.3398W}
{Wang} L.,  {Nitadori} K.,   {Makino} J.,  2020, \mn@doi [\mnras]
  {10.1093/mnras/staa480}, \href
  {https://ui.adsabs.harvard.edu/abs/2020MNRAS.493.3398W} {493, 3398}

\bibitem[\protect\citeauthoryear{{Wang}, {Leigh}, {Liu}  \& {Perna}}{{Wang}
  et~al.}{2021}]{2021MNRAS.505.1053W}
{Wang} Y.-H.,  {Leigh} N. W.~C.,  {Liu} B.,   {Perna} R.,  2021, \mn@doi
  [\mnras] {10.1093/mnras/stab1189}, \href
  {https://ui.adsabs.harvard.edu/abs/2021MNRAS.505.1053W} {505, 1053}

\bibitem[\protect\citeauthoryear{{Yoshida}}{{Yoshida}}{1990}]{1990PhLA..150..262Y}
{Yoshida} H.,  1990, \mn@doi [Physics Letters A]
  {10.1016/0375-9601(90)90092-3}, \href
  {https://ui.adsabs.harvard.edu/abs/1990PhLA..150..262Y} {150, 262}

\bibitem[\protect\citeauthoryear{{Zhu}}{{Zhu}}{2021}]{2021NewA...8501481Z}
{Zhu} Q.,  2021, \mn@doi [\na] {10.1016/j.newast.2020.101481}, \href
  {https://ui.adsabs.harvard.edu/abs/2021NewA...8501481Z} {85, 101481}

\makeatother
\end{thebibliography}




\appendix
\section{Effect of resolution} \label{appendix:resolution}
\begin{figure}
    \centering
    \includegraphics[width=0.5\textwidth]{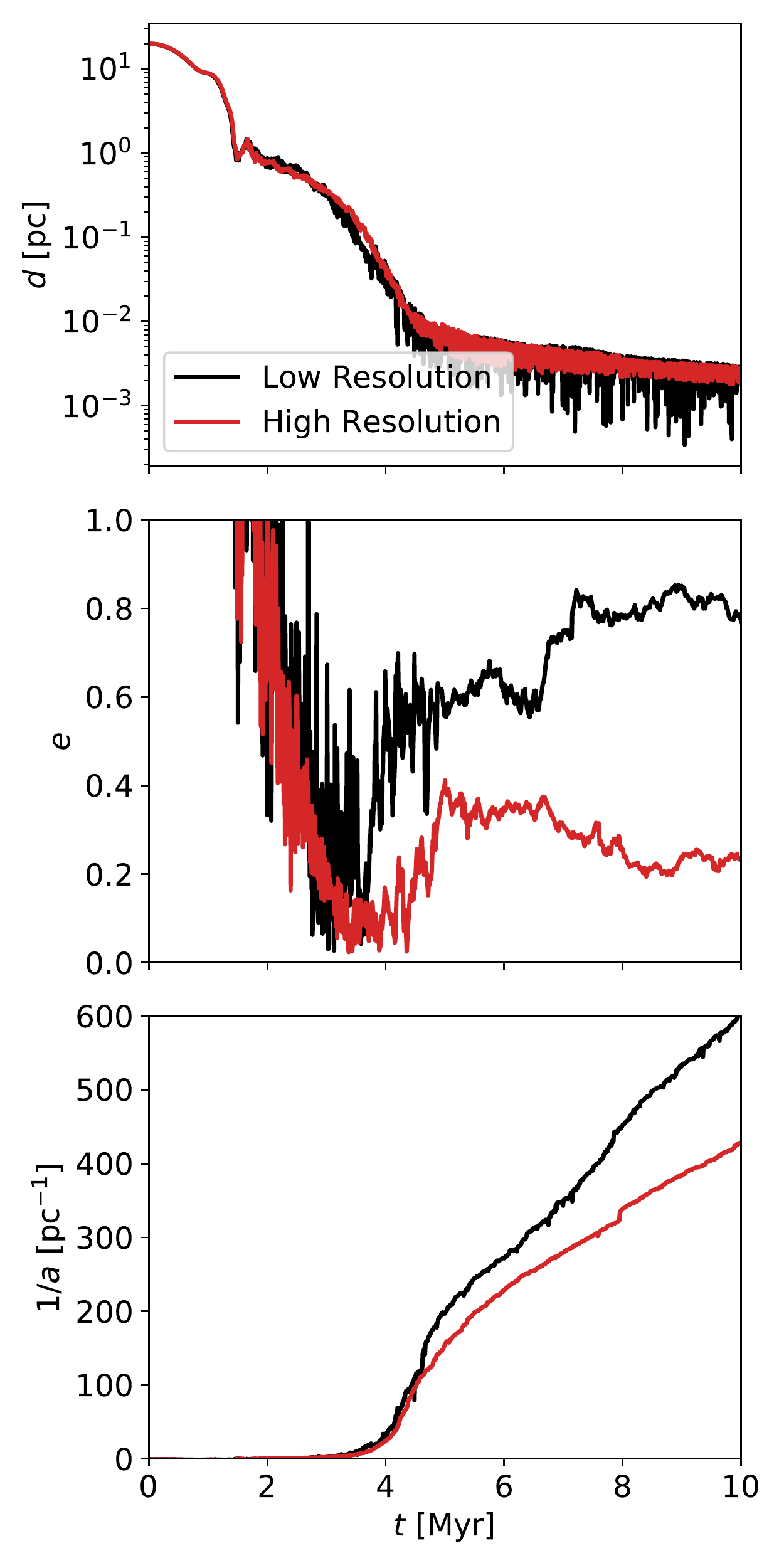}
    \caption{Evolution of the binary parameters for the relaxed $q=0.01$ model presented as a function of time for lower resolution and higher resolution models. We find that although there are no differences in pre-binary phase and the time of binary formation between the lower resolution and higher resolution models, differences appear once the binary is in the bound-binary and the hard binary phases. This is quite notable for the evolution of eccentricity and the rate of hardening where the lower resolution model demonstrates a higher value compared to the high resolution model. The results are in contrast with \citet{Preto2011ApJ...732L..26P} as we find the hardening rate depends on $N$ indicating that the effects of collisional loss-cone refilling cannot be discounted. }
    \label{fig:q001_res_compare}
\end{figure}

\begin{figure}
    \centering
    \includegraphics[width=0.5\textwidth]{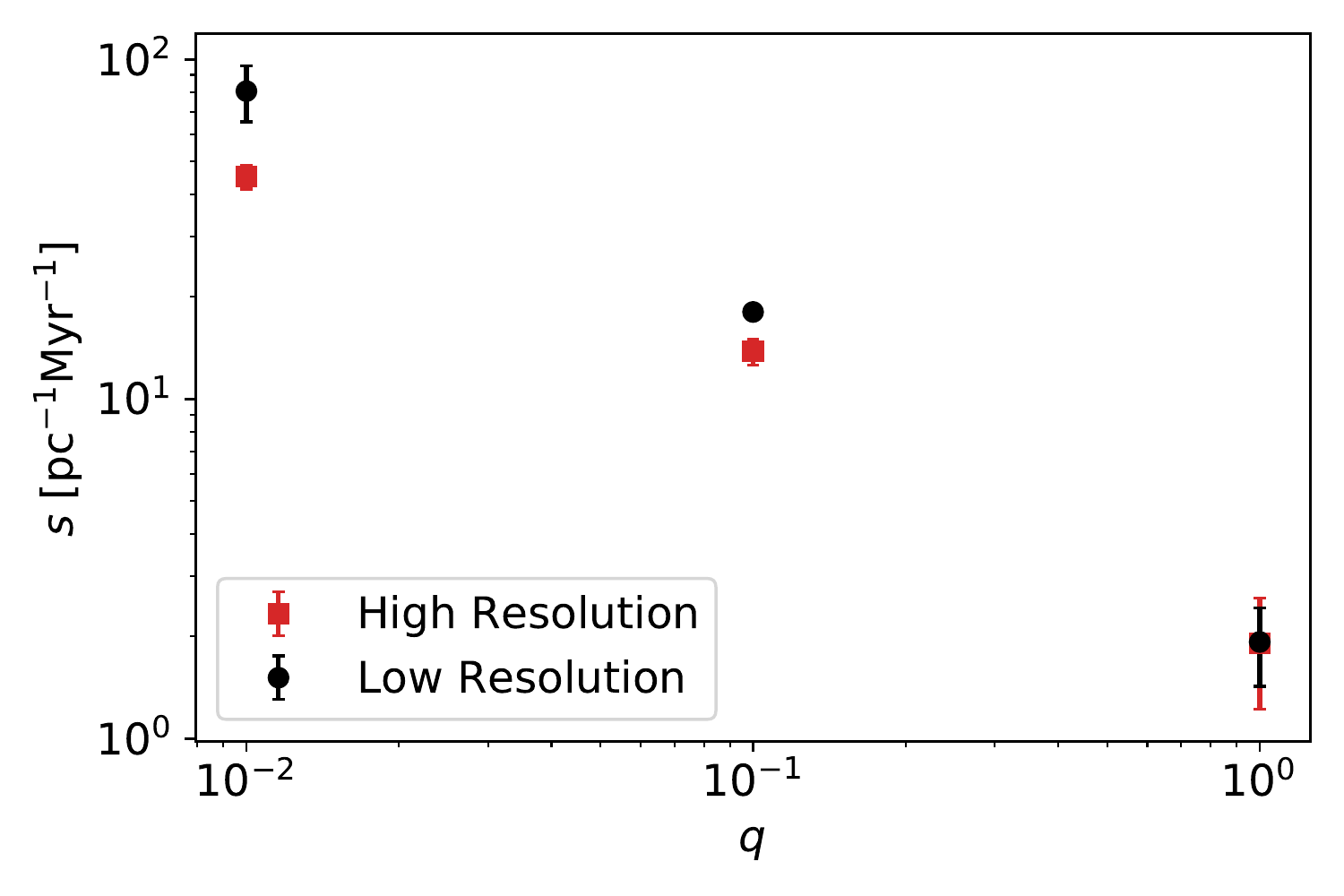}
    \caption{Hardening rates of circular relaxed models with different resolutions presented as a function of the mass-ratio $q$. The hardening rates have been computed by taking the average of the hardening rates every 1 Myr after a hard binary has been formed. The error bars correspond to the standard deviation. We find that the hardening rate strongly depends on $q$ and resolution. As the mass-ratio is lowered, the hardening rate decreases as we increase $N$. Similar observations were noted for the non-relaxed scenario.}
    \label{fig:hard_res_compare}
\end{figure}

To understand the effect of resolution on the results, we simulate lower resolution models of mergers on circular orbits with $N = 4\times 10^{5}$. For clarity, we have only presented the full evolution for the $q=0.01$ models. Looking at Figure \ref{fig:q001_res_compare}, we do not find any differences in the pre-binary phase indicating that the lower resolution simulations are able to simulate the tidal stripping process as accurately as the higher resolution simulations. The time of binary formation is consistent among the two sets of simulations as well. However, once the binary has formed, differences appear between the lower resolution and the higher resolution simulations. In simulations with $N = 4\times 10^{5}$, we find that the binary is able to reach large values of eccentricity which is not observed in the simulations with $N = 1.32 \times 10^{6}$ particles. This was only observed in the $q=0.01$ models but not in the larger mass-ratio models. In addition, as mentioned before, the hardening rate in the lower resolution simulation is about $2\times$ higher than that in the higher resolution simulation. To understand how the resolution plays a role in the hardening rate, we plot the average hardening rate as a function of the mass-ratio in Figure \ref{fig:hard_res_compare}. We find that the hardening rates are consistent among the lower and higher resolution simulations for $q=1.0$. As the mass ratio is lowered, the higher resolution models demonstrate lower rates of hardening. This indicates that although the primary method for loss-cone scattering in the hard binary phase may be collisionless, driven by the non-spherical nature of the merger product, collisional effects cannot be discounted, especially in collisonal systems like NSCs. The dependence of hardening rate on $N$ is more prominent for models with lower $q$. This is in line with the observations of \cite{Vasilev2014ApJ...785..163V} where the authors found that even in non-spherical galaxies, collisional loss cone refilling can play a significant part.

\section{Comparsion between NBODY6 and Taichi}

\begin{figure*}
    
    \centering
    \includegraphics[width=1.0\textwidth]{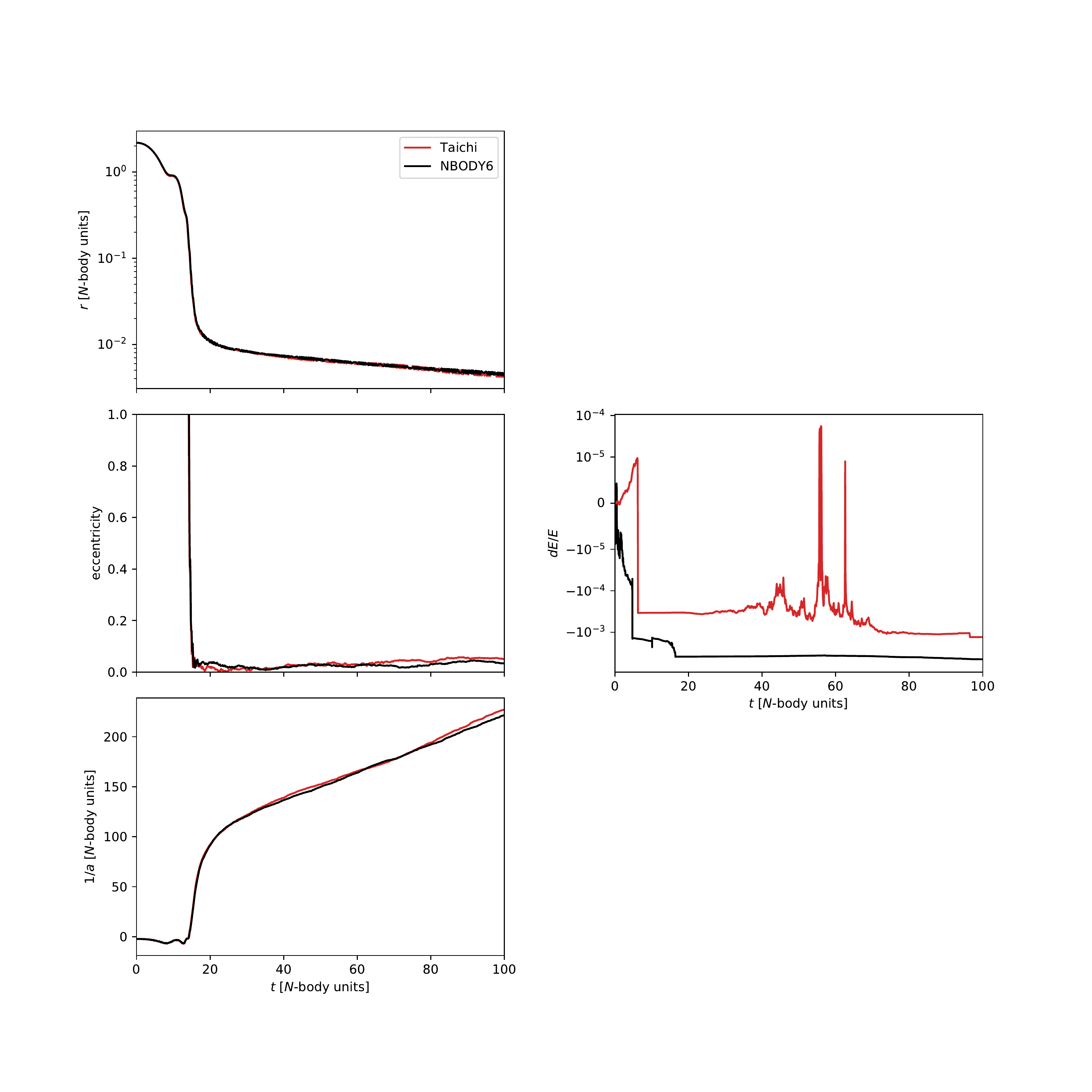}
    \caption{Evolution of the orbital parameters as a function of time for the $q=1.0$ model presented in \citet{2020MNRAS.493.3676O}. Top: Evolution of binary separation. Middle: Evolution of eccentricity. Bottom: Evolution of inverse semi-major axis. Right: Evolution of the relative energy error. We find the results between {\tt\string NBODY6++GPU} and {\tt\string Taichi} are consistent with each other and {\tt\string Taichi} is better at energy conservation by a factor of $\sim 10$ compared to {\tt\string NBODY6++GPU} .}
    \label{fig:nb6_compare}
\end{figure*}

We use the initial conditions for the $q=1.0$ model directly obtained from \cite{2020MNRAS.493.3676O} to test {\tt\string Taichi} against {\tt\string NBODY6++GPU}. We present the evolution of the orbital elements and show that using the parameters chosen in this study, {\tt\string Taichi} is able to simulate the system as accurately as {\tt\string NBODY6++GPU}. Even without the usage of specialized hardware such as GPUs, {\tt\string Taichi} is able to simulate the systems $2\times$ faster than {\tt\string NBODY6++GPU} using 2 GPUs. The energy conservation at the end of the simulation for {\tt\string Taichi} was $\sim 0.1$\% whereas that for {\tt\string NBODY6++GPU} was $\sim 1$\%. For these simulations, the number of particles was the same as that used in \cite{2020MNRAS.493.3676O}, $N=131072$. We find that there are no systemic differences between the evolution of the binaries using the two codes. We find the differences in the evolution of binary separation are minuscule. Even though the eccentricity evolution is inherently stochastic, qualitatively the evolution is similar in both cases. From the plot of the inverse semi-major axis, we deduce that in both scenarios the binary is hardening at similar rates. This indicates that {\tt\string Taichi} is able to model both collisional and collisionless processes as accurately as {\tt\string NBODY6++GPU} and is suitable to handle the class of problems mentioned in this work.


\bsp	
\label{lastpage}
\end{document}